\documentclass[12pt]{iopart}

\usepackage{soul,xcolor}
\setstcolor{red}

\usepackage{graphicx}% Include figure files
\usepackage{dcolumn}% Align table columns on decimal point
\usepackage{bm}% bold math
\usepackage{color}

\usepackage{iopams}

\expandafter\let\csname equation*\endcsname\relax

\expandafter\let\csname endequation*\endcsname\relax

\usepackage{amsmath}
\begin{document}
\def \beq{\begin{equation}}
\def \eeq{\end{equation}}
\def \bea{\begin{eqnarray}}
\def \eea{\end{eqnarray}}
\def \bem{\begin{displaymath}}
\def \eem{\end{displaymath}}
\def \P{\Psi}
\def \Pd{|\Psi(\boldsymbol{r})|}
\def \Pds{|\Psi^{\ast}(\boldsymbol{r})|}
\def \Po{\overline{\Psi}}
\def \bs{\boldsymbol}
\def \bl{\bar{\boldsymbol{l}}}
%\newcommand{\zhm}[1]{{\color{red}#1}}
% \newcommand{\zhm}[1]{{\color{red}#1}}
%  \newcommand{\jtc}[1]{{\color{blue}#1}}

%%%%%%%%%%%%%%%%%%%%%%%%%%%%%%%%%%%%%%%%%%%%%%%%%
\title{Collapse dynamics for two-dimensional space-time nonlocal nonlinear Schr{\"o}dinger equations}
%%%%%%%%%%%%%%%%%%%%%%%%%%%%%%%%%%%%%%%%%%%%%%%%%
\author{Justin T. Cole$^{1^*}$, Abdullah M. Aurko$^2$, and Ziad H. Musslimani$^2$}
%%%%%%%%%%%%%%%%%%%%%%%%%%%%%%%%%%%%%%%%%%%%%%%%%%%%%%%
\address{$^1$  Department of Mathematics, University of Colorado, Colorado Springs, CO 80918, USA}
\address{$^2$  Department of Mathematics, Florida State University, Tallahassee, FL 32306, USA}
\address{*Corresponding author, email: jcole13@uccs.edu}
%%%%%%%%%%%%%%%%%%%%%%%%%%%%%%%%%%%%%%%%%%%%%%%%%%%%%%%%%%%%%%%%%%%%%%%%%%%%%%%%%%%%%%%%%
\begin{abstract}
The question of collapse (blow-up) in finite time is investigated for the two-dimensional (non-integrable) space-time nonlocal 
nonlinear Schr\"odinger equations.
Starting from the two-dimensional extension of the well known AKNS $q,r$ system, 
three different cases are considered: (i) partial and full parity-time (PT) symmetric, 
(ii) reverse-time (RT) symmetric, and (iii) general $q,r$ system. Through extensive numerical experiments, it is shown that collapse of
Gaussian initial conditions depends on the value of its quasi-power. The collapse dynamics (or lack thereof) strongly depends on whether 
the nonlocality is in space or time. A so-called quasi-variance identity is derived and its relationship to blow-up is discussed. Numerical simulations reveal 
 that this quantity reaching zero in finite time does {\it not} (in general) guarantee collapse. An alternative approach to the study of wave collapse is presented 
 via the study of transverse instability of line soliton solutions. In particular, the linear stability problem for perturbed  solitons is formulated for the nonlocal RT 
 and PT symmetric nonlinear Schr\"odinger (NLS) equations. Through a combination of numerical and analytical approaches, the stability spectrum for
 some stationary one soliton solutions is found. Direct numerical simulations agree with the linear stability analysis which predicts filamentation and subsequent blow-up. %collapse. 
\end{abstract}
%%%%%%%%%%%%%%%%%%%%%%%%%%%%%%%%%%%%%%%%%%%%%%%%%%%%%%%%%%%%%%%%%%%%%%%%%%%%%%%%%%%%%%%%%%%

\section{Introduction}
\label{intro}

%%%%%%%%%%%%%%%%%%%%%%%%%%%%%%%%%%%%%%%%%%%%%%%%%%%%
Finite time singularity formation plays a critical role %central role/theme 
in the study of nonlinear evolution equations. In this regard, the major issue at hand is whether or not a smooth initial condition will %would 
collapse or blow-up in finite time. Well-known examples %that have attracted much attention for the last couple of decades 
include fluid type models (e.g., three-dimensional Euler and Navier-Stokes equations), many-body physics and Bose-Einstein condensation (e.g., multi-dimensional Gross-Pitaevskii equation) and nonlinear 
optics/photonics (nonlinear Schr\"odinger (NLS) equations), to name a few \cite{Titi,Kivshar,Robinson,Sackett,Lushnikov,Klein1,Kevrekidis_ODE}. 
While some of the former models are dissipative in nature, the NLS equation is a Hamiltonian %dynamical 
system that also conserves power or density ($L^2$ norm). Physically speaking, in photonic applications, a collapse or blow-up in finite time amounts to a nonlinear self-focusing process \cite{Rasmussen,Fibich1,Fibich2,Fibich3}. This %type of 
instability was originally observed in \cite{Kelley}.
%%%%%%%%%%%%%%%%%%%%%%%%%%%%%%%%%%%%%%%%%%%%%%%%%%%%
%%%%%%%%%%%%%%%%%%%%%%%%%%%%%%%%%%%%%%%%%%%%%%%%%%%%

Another interesting class of nonlinear evolution equations that could exhibit collapse in finite time are the so-called integrable models. They are among the most  studied nonlinear partial differential equations due to their rich mathematical structure and relationship to physical systems.
These equations are in some sense solvable and as a result it is possible to find exact nontrivial solutions, e.g. solitons.
Moreover, many integrable equations are models for physically significant problems, e.g. water waves \cite{Ablowitz_Segur}, electromagnetic waves \cite{Ablowitz}, and plasmas \cite{KP}. Thus, the interplay between integrability and finite time singularity formation is very intriguing.

In 2013, a new integrable equation was introduced by Ablowitz and Musslimani \cite{AM13} given by
\begin{equation}
\label{ablowitz_Musslimani_eqn}
iq_t(x,t) = q_{xx}(x,t) \pm 2q^2(x,t)q^*(-x,t) \;.
\end{equation}
Termed the nonlocal NLS equation, (\ref{ablowitz_Musslimani_eqn}) exhibits a nonlocal parity-time symmetry in the nonlinearity \cite{RKDM2,RKDM3,RKDM4,Guo,Ruter}.
In contrast to the classical NLS, this equation preserves the so-called {\it quasi-power}, rather than the standard power/mass quantity. Through the inverse scattering transform, exact soliton solutions to (\ref{ablowitz_Musslimani_eqn}) were 
found (see Eqs.~(\ref{eq18}) and (\ref{breather_soln}) below) that are localized  in space and periodic in time. In fact, such breather soliton solutions were found to develop a singularity in finite time.

With the introduction of this new class of integrable equations, a large number of works have emerged, primarily focused on finding novel solutions, see for example \cite{MG2,MG3,MG4,Ma3,Yang2}, %,Yang4,Xu,Wen,Ma,Chen1,Li,Yang_Chen,Rao}}, 
and the inverse scattering transform \cite{AblowitzMusslimani3,AblowitzMusslimani4,AblowitzMusslimani5,Ma1,Yang1,Fokas,AblowitzLuoMusslimani1,AblowitzLuoMusslimani2,AblowitzLuoMusslimani3,AblowitzLuoMusslimani4}. On the other hand, studies related to the general question of stability and dynamics of these solutions  have received less attention. 

In this paper, we address the general question of collapse (or blow-up) in finite time for two-dimensional 
(2D) space-time nonlocal NLS systems. In the classical 2D elliptic NLS case with focusing cubic nonlinearity, it is known that collapse is possible whenever the energy (Hamiltonian) of an initial condition is negative \cite{Vlasov,Sulem,Fibich}. In this situation, the variance vanishes in finite time. In addition, it is known that  sufficiently localized initial conditions, whose power is below some fixed threshold value, give rise to solutions that exist globally (for all time) in critical dimensions \cite{Weinstein}.
Many collapse results involve characterizing the blow-up process or dynamics using radially-symmetric ground state solution of the 2D classical cubic  NLS equation \cite{Kuznetsov2,Holmer,Holmer2,Chapman,Killip} and often high power solutions collapse in finite time \cite{Lushnikov2,Aceves}. Additionally, the question of well-posedness in the defocusing NLS equation is also an active area of research \cite{Colliander,Tao}.

%%%%%%%%%%%%%%%%%%%%%%%%%%%%%%%%%%%%%%%%%%%%%%%%%%%%%%%%%%%%%%%%%%%%%%%%%%%%%%%

 An important question that immediately arises is whether or not these results extend to the 2D PT and RT nonlocal NLS systems as well. Direct numerical simulations performed on the PT and RT NLS equations reveal collapse-like behavior for certain classes of initial conditions.  A quasi-variance quantity is introduced whose second-order time derivative is proportional to the (constant) quasi-energy. However, simulations done on the parity-time NLS variant reveal this quantity is not sign-definite, thus making it inadequate for predicting collapse. 
 
In the second part of this work, we study transverse instability of line soliton solutions to the space-time nonlocal NLS systems. Soliton solutions of the cubic NLS are known to be stable in low dimensions \cite{Shatah}. Transverse instability has a long history, dating back to the work of Zakharov and Rubenchik \cite{ZR} on the classical two-dimensional nonlinear Schr\"odinger equation. Since then, the study of transverse instability has been examined in numerous contexts, such as Jacobi elliptic functions \cite{Segur}, dark solitons \cite{Hoefer}, coupled mode systems \cite{Musslimani2},  hyperbolic systems \cite{Deconinck,Deconinck2}, biharmonic operators \cite{Cole2015}, among others. A thorough set of reviews can be found in \cite{Kuznetsov,Kivshar}.
 We investigate the question of collapse in the PT and RT variants through the study of  transverse instability of  line soliton solutions to  the nonlocal NLS-type equations. Through a combination of analytical and numerical means, it is found that line solitons breaks into a periodic train of two-dimensional filaments that ultimately collapse.

The outline of the paper is as follows. In Sec.~\ref{qrsys} we present the 2D nonlocal equations, relevant conserved quantities, and some of their line soliton solutions. A  quasi-variance identity is introduced in Sec.~\ref{gen_var_sec} for the study of wave collapse.  In Sec.~\ref{collpase_dynamics_sec}, extensive numerical simulations highlight the collapse dynamics for Gaussian initial conditions. Subsequently, in Sec.~\ref{linearstab} a linear stability problem, governing the dynamics of the transverse perturbation, is formulated and solved numerically. In Sec.~\ref{asymptotic}, a long wavelength asymptotic theory is developed and shown to agree with numerics. Direct numerical simulations in Sec.~\ref{cases} highlight the full nonlinear evolution of the transverse instability. We conclude in Sec.~\ref{conclude} with remarks for  future work.

%%%%%%%%%%%%%%%%%%%%%%%%%%%%%%%%%%%%%%%%%%%%%%%%%%%%
%%%%%%%%%%%%%%%%%%%%%%%%%%%%%%%%%%%%%%%%%%%%%%%%%%%%
%%%%%%%%%%%%%%%%%%%%%%%%%%%%%%%%%%%%%%%%%%%%%%%%%%%%
\section{Two-dimensional $q,r$ system}
\label{qrsys}
%%%%%%%%%%%%%%%%%%%%%%%%%%%%%%%%%%%%%%%%%%%%%%%%%%%%
%%%%%%%%%%%%%%%%%%%%%%%%%%%%%%%%%%%%%%%%%%%%%%%%%%%%
%%%%%%%%%%%%%%%%%%%%%%%%%%%%%%%%%%%%%%%%%%%%%%%%%%%%
The starting point of this work is the %\zhm{REMOVE:} 
two-dimensional %(2+1)\zhm{D} 
extension of the well-known AKNS $q,r$ system \cite{Ablowitz_Segur}
%%%%%%%%%%%%%%%%%%%%%%%%%%%%%%%%%%%%%%%%%%%%%%%%%%%%
\begin{align}
    \label{eq1}
    iq_t(x,y,t)& = \Delta q(x,y,t) -2r(x,y,t)q^2(x,y,t) , \\
    \label{eq2}
    -ir_t(x,y,t)& = \Delta r(x,y,t) -2q(x,y,t)r^2(x,y,t) ,
\end{align}
%%%%%%%%%%%%%%%%%%%%%%%%%%%%%%%%%%%%%%%%%%%%%%%%%%%%
where $\Delta \equiv \partial_x^2 + \partial_y^2$ and the complex-valued field variables $q,r$ are functions of the two spatial variables $x,y$, and time $t$. 
Here, we refer to $x$ as the longitudinal variable while $y$ the transverse variable. 
In general, these equations are not integrable. The functions $q,r$ are assumed to decay sufficiently fast to zero as $x^2 + y^2 \rightarrow \infty$.
The one-dimensional (1D) reduction (i.e., $q(x,y,t) = q(x,t)$ and $r(x,y,t) = r(x,t)$) was first studied by Ablowitz and Musslimani  in the seminal works \cite{AM13,AblowitzMusslimani3,AblowitzMusslimani4}. Under special symmetry relations, the 1D  
version of Eqs. (\ref{eq1})-(\ref{eq2}) is integrable (possesses a Lax pair) and is exactly solvable by the inverse scattering transform \cite{Ablowitz_Segur}. Equations (\ref{eq1}) and (\ref{eq2}) are conservative  and admit the following  conserved quantities
%%%%%%%%%%%%%%%%%%%%%%%%%%%%%%%%%%%%%%%%%%%%%%%%%%%%
\begin{equation}
    \label{eq3}
    P = \iint  \limits_{\mathbb{R}^2} qr   \hspace{2mm}dx dy , 
\end{equation}
%%%%%%%%%%%%%%%%%%%%%%%%%%%%%%%%%%%%%%%%%%%%%%%%%%%%
%%%%%%%%%%%%%%%%%%%%%%%%%%%%%%%%%%%%%%%%%%%%%%%%%%%%
\begin{equation}
    \label{eq5}
    H = \iint \limits_{\mathbb{R}^2} \left(   q_xr_x + q_yr_y + q^2r^2  \right)  \hspace{2mm} dx dy ,
\end{equation}
%%%%%%%%%%%%%%%%%%%%%%%%%%%%%%%%%%%%%%%%%%%%%%%%%%%%
which we refer to as the quasi-power and quasi-energy, respectively. The classical analog of these quantities is power/mass and Hamiltonian (energy), respectively.  Special relationships between the $q$ and $r$ functions reduce the coupled system (\ref{eq1})-(\ref{eq2}) to a single and often nonlocal equation. In this work, we examine three different types of reductions: one local and two space-time nonlocal. Throughout the rest of the paper, we choose the self-focusing NLS variants. Imposing the relation 
%%%%%%%%%%%%%%%%%%%%%%%%%%%%%%%%%%%%%%%%%%%%%%%%%%%%
\begin{equation}
    \label{eq6}
    r(x,y,t)=- q^{\star}(x,y,t) ,
\end{equation}
%%%%%%%%%%%%%%%%%%%%%%%%%%%%%%%%%%%%%%%%%%%%%%%%%%%%
where $\star$ denotes complex conjugation on system (\ref{eq1}) and (\ref{eq2}), yields the classical  and local (LOC) 2D NLS equation 
%%%%%%%%%%%%%%%%%%%%%%%%%%%%%%%%%%%%%%%%%%%%%%%%%%%%
\begin{equation}
    \label{eq7}
    iq_t(x,y,t)=\Delta q(x,y,t) + 2 q^2(x,y,t) q^\star(x,y,t) .
\end{equation}
%%%%%%%%%%%%%%%%%%%%%%%%%%%%%%%%%%%%%%%%%%%%%%%%%%%%
The classical NLS is a prototypical equation for nonlinear dispersive systems which can be derived in a variety of physical settings \cite{Ablowitz}.  An interesting nonlocal reduction (originally presented in \cite{AblowitzMusslimani4} for the 1D integrable case) is given by
%%%%%%%%%%%%%%%%%%%%%%%%%%%%%%%%%%%%%%%%%%%%%%%%%%%%
\begin{equation}
    \label{eq8}
    r(x,y,t) = - q(x, y,-t),
\end{equation}
%%%%%%%%%%%%%%%%%%%%%%%%%%%%%%%%%%%%%%%%%%%%%%%%%%%%
which, in turn, reduces system (\ref{eq1}) and (\ref{eq2}) to the reverse-time (RT) 2D nonlocal NLS equation
%%%%%%%%%%%%%%%%%%%%%%%%%%%%%%%%%%%%%%%%%%%%%%%%%%%%
\begin{equation}
    \label{eq9}
    iq_t(x,y,t)=\Delta q(x,y,t)  + 2 q^2(x,y,t)q(x, y,-t) .
\end{equation}
%%%%%%%%%%%%%%%%%%%%%%%%%%%%%%%%%%%%%%%%%%%%%%%%%%%%
Notice there is no conjugation in this equation. Another nonlocal reduction (first discovered in \cite{AM13,AblowitzMusslimani3} for the 1D integrable case), is the so-called partial parity-time (PT) reduction, 
%%%%%%%%%%%%%%%%%%%%%%%%%%%%%%%%%%%%%%%%%%%%%%%%%%%%
\begin{equation}
    \label{eq10}
    r(x,y,t)=- q^{\star}(-x,y,t),
\end{equation}
%%%%%%%%%%%%%%%%%%%%%%%%%%%%%%%%%%%%%%%%%%%%%%%%%%%%
which yields 
%%%%%%%%%%%%%%%%%%%%%%%%%%%%%%%%%%%%%%%%%%%%%%%%%%%%
\begin{equation}
    \label{eq11}
    iq_t(x,y,t)=\Delta q(x,y,t) + 2 q^2(x,y,t) q^\star(-x,y,t) .
\end{equation}
%%%%%%%%%%%%%%%%%%%%%%%%%%%%%%%%%%%%%%%%%%%%%%%%%%%%
The nonlinearly induced potential $\mathcal{V}(x,y,t) = 2 q(x,y,t) q^\star(-x,y,t)$ possess the PT symmetry: invariance under $x \rightarrow - x$ and complex conjugation, that is, 
$\mathcal{V}^\star(-x,y,t) = \mathcal{V}(x,y,t)$. 
Additional PT variants can be formulated through the reductions 
%%%%%%%%%%%%%%%%%%%%%%%%%%%%%%%%%%%%%%%%%%%%%%%%%%%%
\begin{align}
   \label{y_nonlocal_sym_a}
    r(x,y,t) & = - q^{\star}(x,-y,t) , \\ 
       \label{y_nonlocal_sym_b}
    r(x,y,t) & = - q^{\star}(-x,-y,t), 
\end{align}
%%%%%%%%%%%%%%%%%%%%%%%%%%%%%%%%%%%%%%%%%%%%%%%%%%%%
which includes nonlocality in the $y$ direction (again partial PT symmetry) and nonlocality in both $x$ and $y$ (which here we refer to as full PT symmetry),  respectively. 
%%%%%%%%%%%%%%%%%%%%%%%%%%%%%%%%%%%%%%%%%%%%%%%%%%%%%%
\subsection{Line solitons}
%%%%%%%%%%%%%%%%%%%%%%%%%%%%%%%%%%%%%%%%%%%%%%%%%%%%%%
In this section, we look for soliton solutions to the general $q,r$ system in Eqs. (\ref{eq1}) and (\ref{eq2}) 
that are homogeneous in $y$ and separable in $x$ and $t$. They are of the form
%%%%%%%%%%%%%%%%%%%%%%%%%%%%%%%%%%%%%%%%%%%%%%%%%%%%
\begin{align} \label{eq12}
q(x,y,t) & =q_0(x)e^{-i\mu t}, \\
 \label{eq13}
r(x,y,t)& =r_0(x)e^{i\mu t},
\end{align}
%%%%%%%%%%%%%%%%%%%%%%%%%%%%%%%%%%%%%%%%%%%%%%%%%%%%
where $q_0(x), r_0(x)$ are assumed to be, in general, complex valued.
Notice that $q$ and $r$ must have temporal phases with opposite signs to completely separate out the time dependence.
It should be noted however that the $q,r$ system admits a larger class of line soliton solutions that are homogeneous in $y$, localized in $x$ and quasi-periodic in time, i.e., 
not in the form given in (\ref{eq12}) and (\ref{eq13}) (see Eq.(\ref{breather_soln}), for example). From (\ref{eq1}) and (\ref{eq2}), we get
%%%%%%%%%%%%%%%%%%%%%%%%%%%%%%%%%%%%%%%%%%%%%%%%%%%%
\begin{equation} \label{eq14}
\mu q_0-q_{0xx}+2r_0q_0^2=0 ,
\end{equation}
%%%%%%%%%%%%%%%%%%%%%%%%%%%%%%%%%%%%%%%%%%%%%%%%%%%%
%%%%%%%%%%%%%%%%%%%%%%%%%%%%%%%%%%%%%%%%%%%%%%%%%%%%
\begin{equation} \label{eq15}
\mu r_0- r_{0xx}+ 2q_0r_0^2=0 .
\end{equation}
%%%%%%%%%%%%%%%%%%%%%%%%%%%%%%%%%%%%%%%%%%%%%%%%%%%%
Special soliton solutions for Eqs. (\ref{eq14}) and (\ref{eq15}) that obey the reductions (\ref{eq6}), (\ref{eq8}), and (\ref{eq10}), respectively, are: 
%%%%%%%%%%%%%%%%%%%%%%%%%%%%%%%%%%%%%%%%%%%%%%%%%%%%
\begin{align}
    \label{eq16}
{\rm LOC:} &  ~~~~~ q_0(x)=\sqrt{\mu} ~ {\rm sech} (\sqrt{\mu}x) , \\ \nonumber \\
    \label{eq17}
{\rm RT:} &  ~~~~~     q_0(x)=\sqrt{\mu} ~ {\rm sech} (\sqrt{\mu}x), \\ \nonumber \\
    \label{eq18}
 {\rm PT:} &  ~~~~~    q_0(x)=\sqrt{\mu} ~ {\rm sech} (\sqrt{\mu}x-i\theta) ,
\end{align}
%%%%%%%%%%%%%%%%%%%%%%%%%%%%%%%%%%%%%%%%%%%%%%%%%%%%
where $\theta$ is a real constant. Note that formula (\ref{eq18}) is undefined at $x= 0$ when $\theta = \pi \left( \frac{1}{2} + n \right)$ for $n \in \mathbb{Z} $. 
 Also,  when $\theta \neq n \pi , n \in \mathbb{Z}$, the soliton solution given in  (\ref{eq18}) is {\it not} a solution to the classical NLS equation given by (\ref{eq7}).  
 Furthermore, note that the PT solution in (\ref{eq18}) {\it is} a solution of RT equation (\ref{eq9}).
Observe that the PT solution in Eq.~(\ref{eq18}) exhibits the symmetry: $q_0^\star(-x) = q_0(x)$. 
As a result, each of these three solutions satisfies the common equation 
%%%%%%%%%%%%%%%%%%%%%%%%%%%%%%%%%%%%%%%%%%%%%%%%%%%%
\begin{equation}
\label{eqstat}
\mu q_0(x)-q_{0xx}(x)-2q_0^3(x)=0 .
\end{equation}
%%%%%%%%%%%%%%%%%%%%%%%%%%%%%%%%%%%%%%%%%%%%%%%%%%%%
%%%%%%%%%%%%%%%%%%%%%%%%%%%%%%%%%%%%%%%%%%%%%%%%%%%
\section{Quasi-variance identity}
\label{gen_var_sec}
%%%%%%%%%%%%%%%%%%%%%%%%%%%%%%%%%%%%%%%%%%%%%%%%%%%
In this and the following sections, collapse (or blow-up) in finite time of the nonlocal NLS equations and, more generally, the $q,r$ system is studied. 
One of the main findings is that well-known collapse results from the classical NLS theory do not necessarily extend to the nonlocal systems. To see that, we introduce the quasi-variance 
%%%%%%%%%%%%%%%%%%%%%%%%%%%%%%%%%%%%%%%%%%%%%%%%%%%
\begin{equation}
    \label{eq53}
    V(t)= \iint \limits_{\mathbb{R}^2} |{\bf x} |^2 q(x,y,t) r(x,y,t)  \hspace{2mm}dx dy ,
\end{equation}
%%%%%%%%%%%%%%%%%%%%%%%%%%%%%%%%%%%%%%%%%%%%%%%%%%%
where ${\bf x}  \equiv (x , y)$. The study of this quantity is  motivated by a natural generalization of the second moment
%%%%%%%%%%%%%%%%%%%%%%%%%%%%%%%%%%%%%%%%%%%%%%%%%%%
\begin{equation}
    \label{eq52}
    \widetilde{V}(t)= \iint \limits_{\mathbb{R}^2} |{\bf x} |^2 |q|^2 dx dy ,
\end{equation}
%%%%%%%%%%%%%%%%%%%%%%%%%%%%%%%%%%%%%%%%%%%%%%%%%%%
which can be used to measure the localization extent of the functions $r$ and $q$. For the LOC  reduction, $r(x,y,t) = - q^{\star}(x,y,t)$,  these two quantities are directly related via $\widetilde{V}(t) = - V(t)$. %\zhm{DO WE NEED THIS:}This generalized variance is discussed in more detail below. 
%\zhm{REMOVE}Consider functions $q,r$ that decay sufficiently fast as $x^2 + y^2 \rightarrow \infty$.
Differentiating $V(t)$ %\zhm{NO NEED}in Eq.~(\ref{eq53}) 
with respect to time, we obtain
%%%%%%%%%%%%%%%%%%%%%%%%%%%%%%%%%%%%%%%%%%%%%%%%%%%
\begin{equation}
    \label{eq54}
    \frac{dV}{dt}= 2 i \iint \limits_{\mathbb{R}^2} {\bf x} \cdot \left( r  \nabla q  - q \nabla r \right) dx dy ,
\end{equation}
%%%%%%%%%%%%%%%%%%%%%%%%%%%%%%%%%%%%%%%%%%%%%%%%%%%
where $\nabla \equiv (\partial_x , \partial_y)$. Differentiating Eq.~(\ref{eq54}) with respect to $t$ gives
%%%%%%%%%%%%%%%%%%%%%%%%%%%%%%%%%%%%%%%%%%%%%%%%%%%
\begin{equation}
    \label{eq55}
    \frac{d^2V}{dt^2}= 8 H ,
\end{equation}
%%%%%%%%%%%%%%%%%%%%%%%%%%%%%%%%%%%%%%%%%%%%%%%%%%%
where $H$ is the conserved (constant) quasi-energy given in Eq.~(\ref{eq5}). Integrating (\ref{eq55}) leads to
%%%%%%%%%%%%%%%%%%%%%%%%%%%%%%%%%%%%%%%%%%%%%%%%%%%
\begin{equation}
    \label{eq56}
     V(t) = 4 H t ^2 +  V'(0)  t+V(0) .
\end{equation}
%%%%%%%%%%%%%%%%%%%%%%%%%%%%%%%%%%%%%%%%%%%%%%%%%%%
The %This 
quasi-variance (\ref{eq53}) or (\ref{eq56}) has the following important properties: %\zhm{REMOVE}listed below
%%%%%%%%%%%%%%%%%%%%%%%%%%%%%%%%%%%%%%%%%%%%%%%%%%%%
%%%%%%%%%%%%%%%%%%%%
\begin{enumerate}
\item  LOC, $r(x,y,t) = - q^{\star}(x,y,t)$:   
%%%%%%%%%%%%%%%%%%%%%
 \begin{itemize}
\item  $V(t) \le 0$ 
\end{itemize}
%%%%%%%%%%%%%%%%%%%%
\item RT, $r(x,y,t) = - q(x,y,-t)$: 
%%%%%%%%%%%%%%%%%%%%
 \begin{itemize}
\item  $V(t) = V(-t) $
\item if $q(x,y,0)$ is real, then $V(t), H \in \mathbb{R}$, $V'(0) = 0$ 
\end{itemize}
%%%%%%%%%%%%%%%%%%%%
\item PT, $r(x,y,t) = - q^{\star}(-x,y,t)$: 
%%%%%%%%%%%%%%%%%%%%
 \begin{itemize}
\item  $V(t), H, V'(0) \in \mathbb{R} $
\end{itemize}
%%%%%%%%%%%%%%%%%%%%
\item General $q,r$ system: 
%%%%%%%%%%%%%%%%%%%%
 \begin{itemize}
\item  Suppose $q(x,y,0)$ and $r(x,y,0)$ are real-valued, such that $V'(0) \in \mathbb{R}$. Then $V(t), H \in \mathbb{R}$
\end{itemize}
%%%%%%%%%%%%%%%%%%%%
\end{enumerate}
%%%%%%%%%%%%
For most cases considered below, we choose initial conditions that lead to $V$ and $H$ being real.
From Eq.~(\ref{eq56}), it is clear that when ${\rm sgn} \left( H \right) = - {\rm sgn} \left( V(0) \right)$, there exists a finite time $t = t_* < \infty $ when %such that 
$V(t_*) = 0$. The implication of this result for the LOC case  is clear: it predicts collapse in finite time \cite{Sulem,Fibich}.
However, for the PT and RT reductions, the following scenarios are possible: (a) $V(t_*) = 0$, 
without collapse; (b) $V(t_*) = 0$, yet collapse could occurs for $t > t_*$. 
%%%%%%%%%%%%%%%%%%%%%%%%%%%%%%%%%%%%%%%%%%%%%%%%%%%
\section{Collapse dynamics of Gaussian initial conditions}
\label{collpase_dynamics_sec}
%%%%%%%%%%%%%%%%%%%%%%%%%%%%%%%%%%%%%%%%%%%%%%%%%%%
In this section we aim to study collapse properties associated with the two-dimensional $q,r$ system (\ref{eq1})-(\ref{eq2}). For simplicity, we consider Gaussian initial conditions 
%%%%%%%%%%%%%%%%%%%%%%%%%%%%%%%%%%%%%%%%%%%%%%%%%%%
\begin{equation}
\label{gaussian}
q(x,y,0)  = A e^{- B(x-x_0)^2 - C (y - y_0)^2 -ivx}  , 
\end{equation}
%%%%%%%%%%%%%%%%%%%%%%%%%%%%%%%%%%%%%%%%%%%%%%%%%%%
where $A,B,C >0$ and $x_0, y_0, v  \in \mathbb{R}$. Unlike the classical NLS equation, where power and Hamiltonian are independent of  the shifts $x_0, y_0$, here (for both PT and RT) these parameters affect the quasi-power and quasi-energy.
As we shall see, the collapse dynamics profoundly depend on the displacement and velocity of the initial condition. We point out that $v \not=0$ is important for distinguishing between the RT and LOC cases. In fact, when $v = 0$ the two conserved quantities, i.e. quasi-power and quasi-energy, reduce to their classical counterparts, i.e. power and Hamiltonian. In this case (\text{RT}-NLS), $v=0$ reduces back to the classical (local) NLS.
To track the blow-up dynamics, we monitor the time evolution of the maximum amplitudes of the functions $q,r$ defined by 
%%%%%%%%%%%%%%%%%%%%%%%%%%%%%%%%%%%%%%%%%%%%%%%%%
\begin{align}
\label{eqqmax}
q_{\max}(t)&  = \max_{(x,y) \in \mathbb{R}^2} |q(x,y,t)|, \\ \nonumber
 r_{\max}(t) &= \max_{(x,y) \in \mathbb{R}^2} |r(x,y,t)|.
\end{align}
%%%%%%%%%%%%%%%%%%
%where $\Omega$ is the computational domain. 
%%%%%%%%%%%%%%%%%%%%%%%%%%%%%%%%%%%%%%%%%%%%%%
Before presenting our results, we first describe the computational methods used to obtain all numerical findings. To this end, we denote by $ \mathcal{F},  \mathcal{F}^{-1}$ the two-dimensional forward and inverse Fourier transforms defined (for suitable functions) by 
%%%%%%%%%%%%%%%%%%%%%%%%%%%%%%%%%%%%%%%%%
\begin{equation}
\mathcal{F}(q)\equiv\widehat{q}({\bf k},t) 
= 
\iint  \limits_{\mathbb{R}^2}  q({\bf x},t) e^{-i {\bf k} \cdot {\bf x}} dx dy \;,
\end{equation}
%%%%%%%%%%%%%%%%%%%%%%%%%%%%%%%%%%%%%%%%%
\begin{equation}
q({\bf x},t)  \equiv \mathcal{F}^{-1}(\widehat{q}) 
= \frac{1}{4\pi^2}
\iint  \limits_{\mathbb{R}^2}  \widehat{q}({\bf k},t) e^{i {\bf k} \cdot {\bf x}} dk_x dk_y \;,
\end{equation}
%%%%%%%%%%%%%%%%%%%%%%%%%%%%%%%%%%%%%%%%%
where ${\bf k} = (k_x,k_y)$ is the 2D Fourier wavevector. 
Taking the Fourier transform of the $q,r$ system (\ref{eq1})-(\ref{eq2}) we obtain
%%%%%%%%%%%%%%%%%%%%%%%%%%%%%%%%%%
\begin{align}
    \label{q_fourier}
    i \widehat{q}_t& = - |{\bf k}|^2 \widehat{q} -2 \widehat{\left(rq^2\right)} , \\
    \label{r_fourier} 
    -i \widehat{r}_t& = - |{\bf k}|^2 \widehat{r} -2\widehat{\left(qr^2\right)} .
\end{align}
%%%%%%%%%%%%%%%%%%%%%%%%%%%%%%%%%%
The integrating factor phase transformations 
%%%%%%%%%%%%%%%%%%%%%%%%%%%%%%%%%%
\begin{align}
    \label{qr_fourier2a}
\widehat{q}({\bf k},t) = \widehat{\xi}({\bf k},t) e^{i |{\bf k}|^2 t}, ~~~~~
\widehat{r}({\bf k},t) = \widehat{\rho}({\bf k},t) e^{-i |{\bf k}|^2 t},
\end{align}
 yield the coupled system of differential equations
%%%%%%%%%%%%%%%%%%%%%%%%%%%%%%%%%%
\begin{align}
    \label{qr_fourier2}
     \widehat{\xi}_t =  2i e^{-i |{\bf k}|^2 t} \widehat{\left(rq^2\right)} , ~~~~~
    \widehat{\rho}_t = -2i e^{i |{\bf k}|^2 t} \widehat{\left(qr^2\right)} ,
\end{align}
%%%%%%%%%%%%%%%%%%%%%%%%%%%%%%%%%%
where $q({\bf x}, t) = \mathcal{F}^{-1}\left[ \widehat{\xi}({\bf k}, t) e^{i |{\bf k}|^2 t}  \right]$ and 
$r({\bf x}, t) = \mathcal{F}^{-1}\left[ \widehat{\rho}({\bf k}, t) e^{-i |{\bf k}|^2 t}  \right]$. System (\ref{qr_fourier2}) is numerically integrated using a pseudo-spectral method (in space) and a fourth-order Runge-Kutta (RK4) time stepping scheme \cite{Yang}. Spatial derivatives were approximated by the discrete Fourier transform on a large computational domain. Additionally, system (\ref{qr_fourier2}) is subject to the following initial conditions: $q(x,y,0)$ is given by (\ref{gaussian}) and

%%%%%%%%%%%%%%%%%%%%%%%%%%%%%%%%%%%%%%%%%%%%%%%%%%%%
\begin{align}
    \label{eq16a1}
\!\!\!{\rm LOC:} &  ~~~ r({\bf x},0)  =  - A e^{- B(x-x_0)^2 - C (y - y_0)^2 +ivx}, \;v=0, \\ \nonumber \\
    \label{eq17a2}
{\rm RT:} &  ~~~     r({\bf x},0)  =  - A e^{- B(x-x_0)^2 - C (y - y_0)^2 -ivx}, \; v=\pi/x_0, \\ \nonumber \\
    \label{eq18a3}
 {\rm Partial ~ PT:} &  ~~~    r({\bf x},0)  =  - A e^{- B(x+x_0)^2 - C (y - y_0)^2 - ivx}, \; v=0,  \\ \nonumber \\
 \label{full_PT_r}
 {\rm Full ~ PT:} &  ~~~    r({\bf x},0)  =  - A e^{- B(x+x_0)^2 - C (y + y_0)^2 - ivx}, \; v=0.  
 \end{align}
%%%%%%%%%%%%%%%%%%%%%%%%%%%%%%%%%%%%%%%%%%%%%%%%%%%%

%%%%%%%%%%%%%%%%%%%%%%%%%%%%%%%%%%%%%%%%%%%%%%%%%%
\begin{figure}
     \centering
     \includegraphics[scale=0.45]{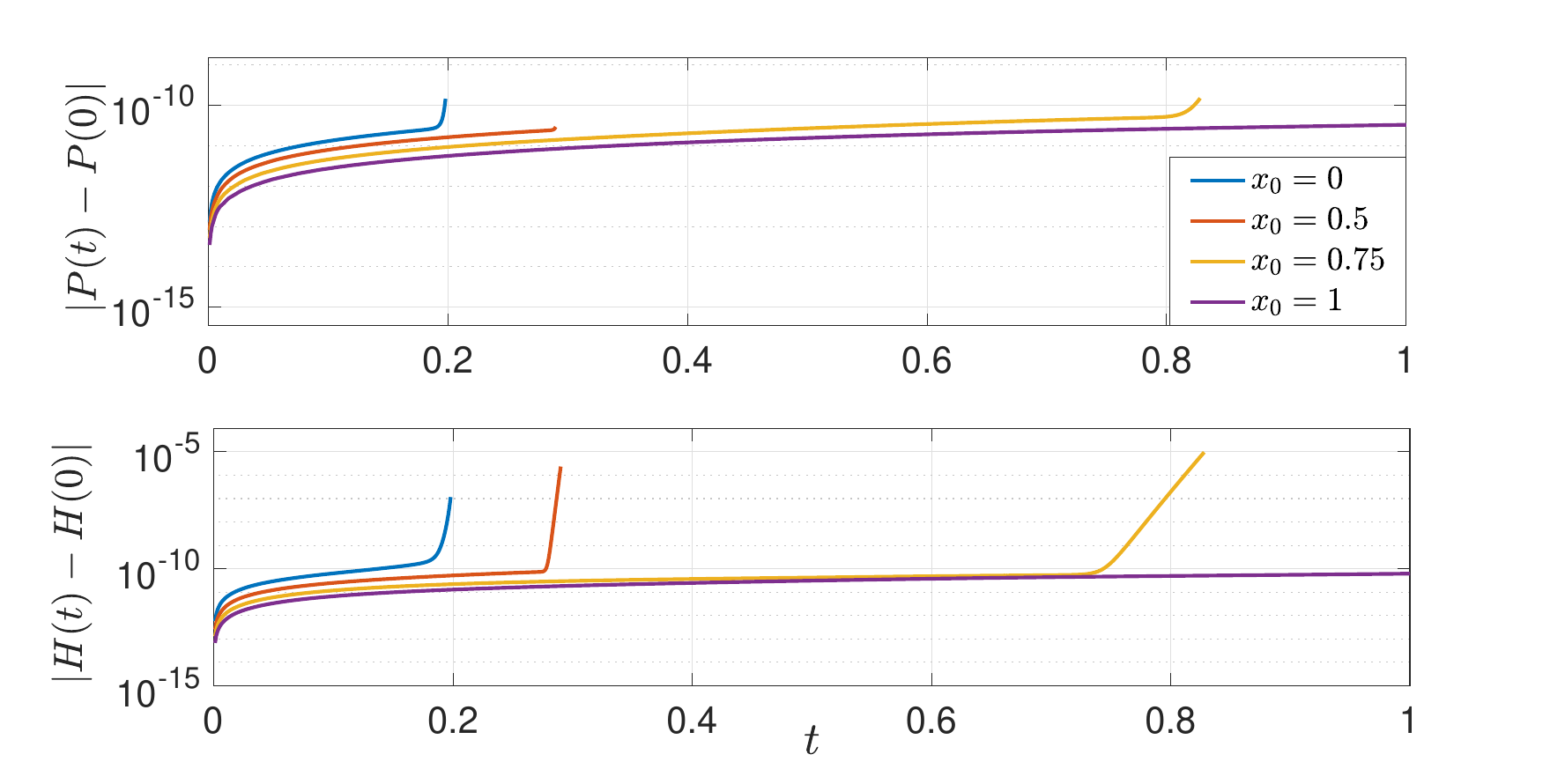}
\caption{The numerical error in the quasi-power (\ref{eq3}) and quasi-energy (\ref{eq5}) for the RK4 Fourier integrating factor method. The integrals appearing in (\ref{eq3}) and (\ref{eq5}) were evaluated spectrally. These curves correspond to the partial PT symmetric case ($r(x,y,t) = -q^*(-x,y,t)$) with  initial conditions and parameters  identical to those in Fig.~\ref{PT_collapse_summary}.} %Note that $x_0=0,0.5,0,75$ are collapse cases, while 
%$x_0 = 1$ \zhm{is not}.}}
        \label{energy_conserve_manuscript}
\end{figure}
%%%%%%%%%%%%%%%%%%%%%%%%%%%%%%%%%%%%%%%%%%%%%%%%%%%
All numerical results obtained for the collapse dynamics (up until collapse starts to develop) are performed on a square computational window $[-15, 15] \times [-15,15]$. We took $N = 2048$ grid points in both $x$ and $y$ directions with grid spacing $\delta x = \delta y \approx 0.0146 $ and time step 
$\delta t = 10^{-5}$. Simulations are performed over the numerical time interval $0 \le t \le 2$.
Such choice was observed to effectively maintain zero boundary conditions up to the collapse time. To gauge the  accuracy of our numerical scheme, we show in Fig.~\ref{energy_conserve_manuscript} the numerically computed conserved quantities (\ref{eq3}) and (\ref{eq5}) for the partial PT case as a function of $t$. In the absence of collapse, these conserved quantities remained accurate (relative to their initial values) up to ten decimal places until reaching the singularity time. As a result, our numerical method yield an accurate depiction of the solution behavior valid up to the onset of collapse. More sophisticated methods are needed to carefully probe the dynamics near and at the singularity. Our goal here is to give preliminary indications of the collapse behavior.
%a future work will implement more precise numerical simulations.
%The number of grid points and time steps are carefully chosen to resolve the singularity formation. We monitored the time evolution of the quasi-power and quasi-energy as well as the relevant symmetries, i.e. (\ref{eq8}) and (\ref{eq10}), to verify they are indeed conserved.
%%%%%%%%%%%%%%%%%%%%%%%%%%%%%%%%%%%%%%%%%%%%%%%%%%%%%%%%%%%%%%%%%%%%%%%%%%%%%%%%%
\subsection{RT collapse dynamics}
\begin{figure}
     \centering
     \includegraphics[scale=0.625]{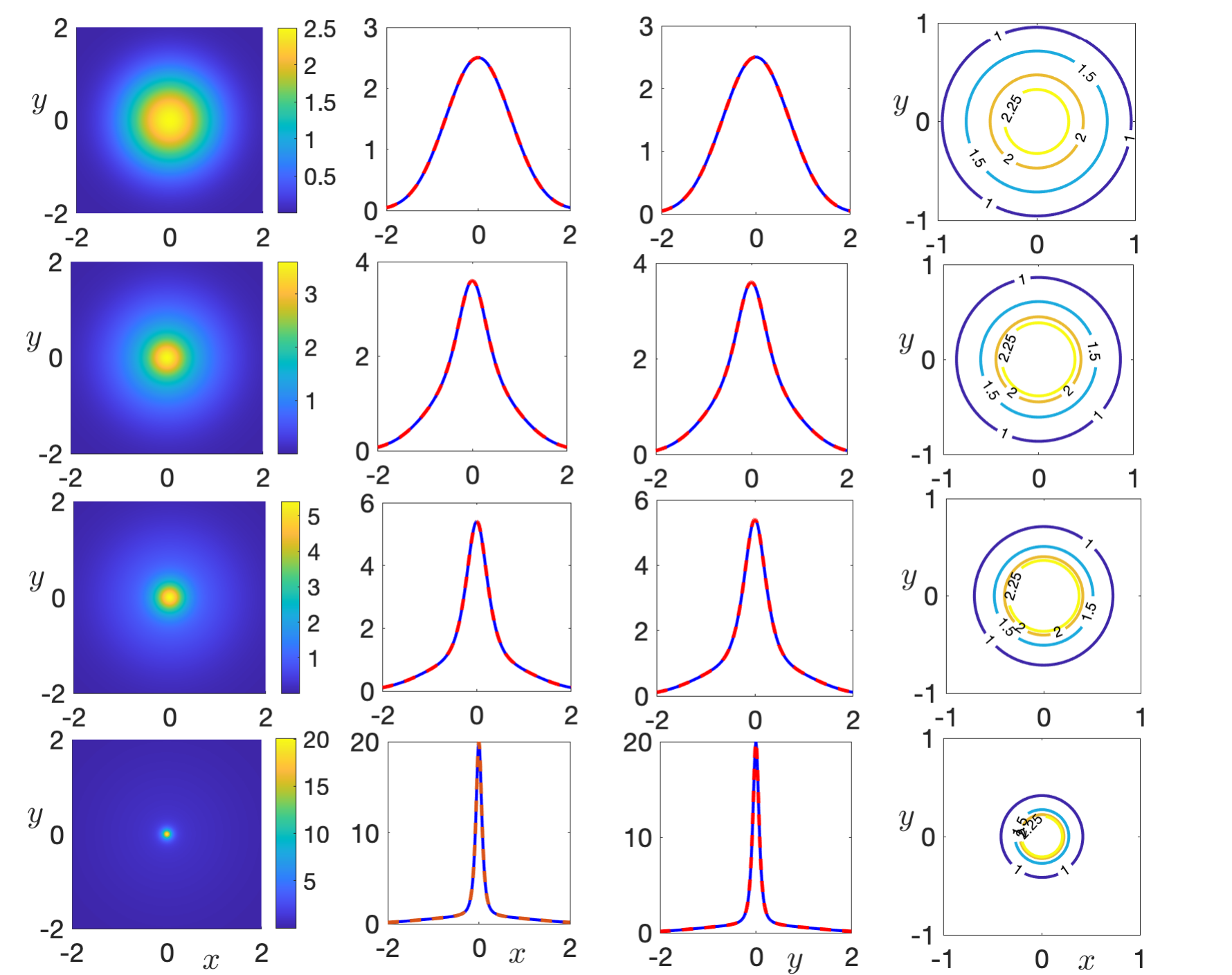}
\caption{Snapshots of the collapsing profile for the LOC NLS system with initial condition given in (\ref{gaussian}) and (\ref{eq16a1}) with $(x_0,y_0) = (0,0)$ and $A = 2.5,B=C=1$. Time evolves from top-to-bottom. Rows correspond to $t= 0, 0.1, 0.15, 0.198$, respectively. First left most column:  top views of $|q(x,y,t)|$ 
and $|r(x,y,t)|$. Second column: $y = 0$ profile slice, $|q(x,0,t)|$ (solid) and $|r(x,0,t)|$ (dashed). Third column: $x = 0$ profile cut, $|q(0,y,t)|$ (solid) and $|r(0,y,t)|$ (dashed). Fourth column: level curve contours denoting fixed magnitudes $|q(x,y,t)| = c$
with values of $c$: 1, 1.5, 2, 2.25. The level curves for $|r(x,y,t) |$ are identical.}  
        \label{profile_summary_NLS}
\end{figure}

In this section we study collapse dynamics associated with the RT-NLS Eq.~(\ref{eq9}) subject to the Gaussian initial data given in (\ref{gaussian}) with 
velocity $v = n \pi / x_0, n \in \mathbb{Z}$. This choice of velocity is made to guarantee the reality of both the quasi-power and quasi-energy (but not necessarily the quasi-variance).
The RT symmetry (like all others) is imposed through the condition (\ref{eq8}) at t=0. Notice that a completely real initial condition ($v=0$ in our case) will reduce to the LOC equation (\ref{eq6}). Hence, later in this section we consider complex initial condition corresponding to $v\ne 0$.
The initial quasi-variance (\ref{eq53}), quasi-power (\ref{eq3}), and quasi-energy (\ref{eq5}), 
are given by (recall that $r(x,y,0) = - q(x,y,0)$): \\\\
%%%%%%%%%%%
{\bf RT:} 
%%%%%%%%%%%%%%%%%%%%%%%%%%%%%%%%%%%%%%%%%%%%%%%%%%%
\begin{align}
\label{V_NLS}
&V(0)  = \frac{\pi A^2 \left[ - B^2 x_0^2 + C \left( n^2 \pi^2 + 4i \pi n B x_0^2 - B x_0^2 \left(1 + 4 B [x_0^2 + y_0^2] \right) \right) \right] }{8 \sqrt{B^5 C^3} x_0^2} \exp\left( - \frac{ n^2 \pi^2 }{2 B x_0^2} \right) , \\
\label{g1_NLS}
&P  = - \frac{\pi A^2}{2 \sqrt{BC}} \exp\left( - \frac{ n^2 \pi^2 }{2 B x_0^2} \right) , \\
\label{g3_NLS}
&H  =- \frac{\pi A^2 (B+C) }{2 \sqrt{BC}} \exp\left( - \frac{n^2 \pi^2}{2B x_0^2} \right) +  \frac{\pi A^4 }{4 \sqrt{BC}} \exp\left( - \frac{n^2 \pi^2}{B x_0^2} \right) .
\end{align}
%%%%%%%%%%%%%%%%%%%%%%%%%%%%%%%%%%%%%%%%%%%%%%%%%%%
Observe that the quasi-power in (\ref{g1_NLS})  is negative-definite. This is due to the minus sign that appears in the RT reduction (\ref{eq8}).
First, consider the stationary (zero velocity) case when $n= 0$. Here, all the above quantities are real-valued. Observe that  ${\rm sgn} \left( V(0)  \right)= -1$. Moreover,  we see that $|V(0)|$  is 
indirectly related to the initial width of the Gaussian at a fixed initial location $x_0, y_0$. As such, when the Gaussian becomes more narrow ($B,C \rightarrow \infty$) the quantity $V(0) \rightarrow 0^{-}$. On the other hand, as $B,C \rightarrow 0$ then $V(0) \rightarrow - \infty$.
%%%%%%%%%%%%%%%%%%%%%%%%%%%%%%%%%%%%%%%%%%%%%%%%%%%
As a result, the quasi-variance is guaranteed to go to zero in finite time for Gaussian initial conditions (\ref{gaussian}) when $ H >0$, or
%%%%%%%%%%%%%%%%%%%%%%%%%%%%%%%%%%%%%%%%%%%%%%%%%%%
\begin{equation}
\label{NLS_collapse_cond}
{\rm \bf{RT}}: ~~~   B +C  <   \frac{A^2}{2}  , ~~~~ n = 0 .
\end{equation}
%%%%%%%%%%%%%%%%%%%%%%%%%%%%%%%%%%%%%%%%%%%%%%%%%%%
 Note that this can only be satisfied in the self-focusing case and these results are independent of the initial position, $x_0$ and $y_0$.  
 Physically, this case corresponds to Gaussians that have large amplitudes  relative to their cross-sections. 
 Inequality (\ref{NLS_collapse_cond}) is a sufficient condition for guaranteeing collapse in finite time of the LOC NLS equation when $H > 0$ \cite{Fibich}, since $V(0) < 0$. 
 %%%%%%%%%%%%%%%%%%%%%%%%%%%%%%%%%%%%%%%%%%%%%%%%%
 Our numerical simulations confirm these results. 
 A typical blowup for the LOC NLS equation ($v=0$) is shown in Fig.~\ref{profile_summary_NLS}. As expected, both the $q$ and $r$ fields maintain a radially-symmetric profile as they collapse.  When (\ref{NLS_collapse_cond}) is satisfied, the maximum amplitude given in (\ref{eqqmax}) grows unboundedly and the width of the wavefunction $q$ appears to be narrowing.  Furthermore, the rightmost column in Fig.~\ref{profile_summary_NLS} shows the level curve rings contracting as the blow-up develops, which agrees with the predicted self-similar blow-up profile \cite{Fibich,Merle2004}. When inequality (\ref{NLS_collapse_cond}) is violated, the solutions can gradually diffract. 
 
 %%%%%%%%%%%%%%%%%%%%%%%%%%%%%%%%%%%%%%%%%%%%%%%%%%
%%%%%%%%%%%%%%%%%%%%%%%%%%%%%%%%%%%%%%%%%%%%%%%%%%%
 %%%%%%%%%%%%%%%%%%%%%%%%%%%%%%%%%%%%%%%%%%%%%%%%%%%%%%%%%%%%%%%%%%%%%%%%%%%%%%%%%%%%%
 
 Next, let us consider the non-stationary case corresponding to $n  = 1$ ($v \not= 0$). Here the RT NLS equation is genuinely different 
 from its classical (LOC) counterpart since no complex conjugation is present.  For the choice of $v =  \pi / x_0 ,$  the quasi-power (\ref{g1_NLS}) and quasi-energy (\ref{g3_NLS}) are completely real, but the variance (\ref{V_NLS}) is complex. Notice that the quasi-power in (\ref{g1_NLS}) is directly related to $x_0^2$. 
 That is, as $x_0^2$ increases, $P$ decreases (becomes more negative). This is profoundly different from the LOC case for which the power $P$ (or the $L^2$ norm) is independent of $x_0.$
 %%%%%%%%%%%%%%%%%%%%%%%%%%%%%%%%%%%%%%%%%%%%%%%%
We numerically measure the (approximate) time $t_s$ for the singularity to form as a function of the Gaussian parameter $x_0$ or  $P$. More precisely, we define the (nearly) singular time as
 %%%%%%%%%%%%%%
 \begin{equation}
 \label{singularity_time_define}
 t = t_s ~~{\rm when}~~ q_{\max}(t_s) = q_c ,
 \end{equation}
  %%%%%%%%%%%%%%
 (see (\ref{eqqmax})) for some $q_c \gg 1$, where $q_{\max}(0) \ll q_c$. As $t\rightarrow t_s^{-}$, the maximum amplitude of the solution, its $L^2$ norm ($||q||_2$), as well as 
 $|| \nabla q||_2$ approach ``infinity", i.e., can grow unboundedly.  In all simulations reported below, we took $q_c = 20$.  The choice of this value for $q_c$  is  arbitrary, but based on extensive numerical experiments seems to be sufficient enough to capture the singularity time when the solution collapses. Moreover, near the collapse time, our numerical accuracy starts to decline (see Fig.~\ref{energy_conserve_manuscript}). Simulations are performed over the numerical time interval $0 \le t \le 2$.
  
\begin{figure}
     \centering
     \includegraphics[scale=0.6]{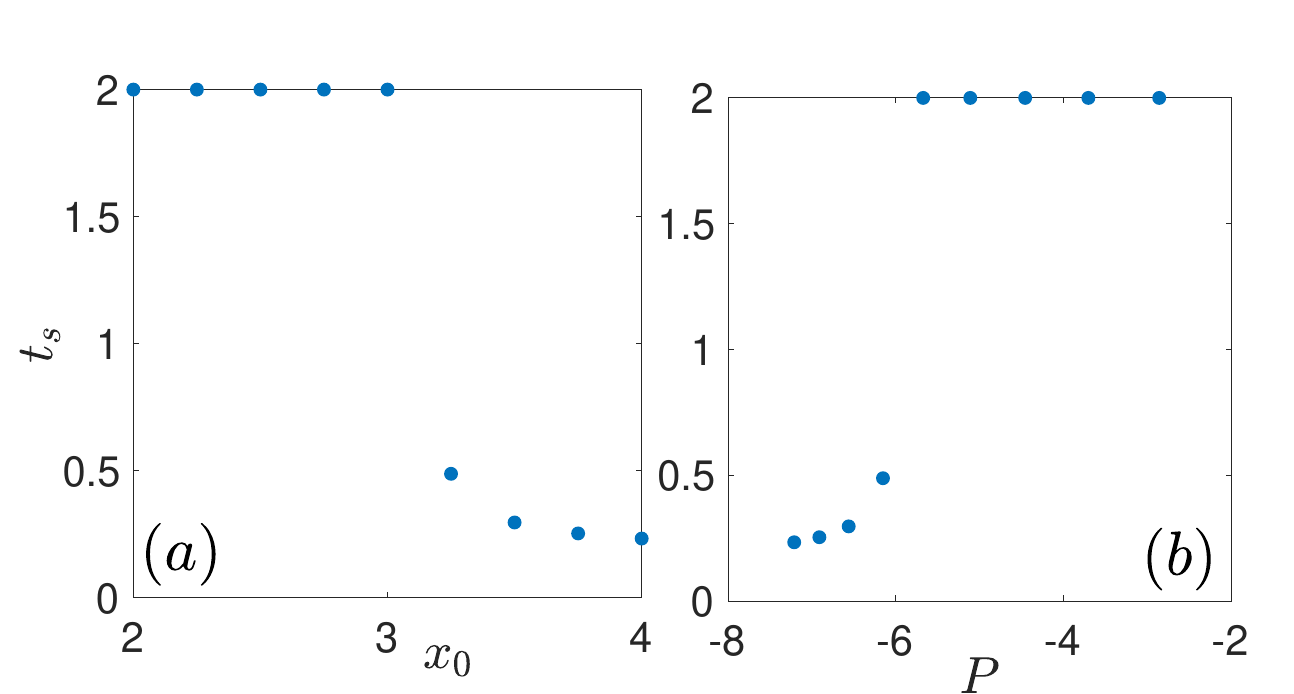}
\caption{The singularity time (\ref{singularity_time_define}) for RT Gaussian initial data (\ref{gaussian}) with $A = 2.5, B = C = 1, y_0 = 0.5, v = \pi / x_0, n = 1$, and different values of (a) $x_0$ and (b) $P$. Blue dots are numerical data.} 
        \label{RT_collpase_summary}
\end{figure}

  Results that indicate blow-up are shown in Fig.~\ref{RT_collpase_summary}. For relatively low quasi-powers ($P \ll -1$), the peak amplitude appears to grow unbounded. Physically, this is associated with moving the Gaussian centers away from the origin, or equivalently, slow velocity $v = \pi/x_0$. On the other hand, when the Gaussian centers are near the origin ($|x_0|\ll1$), the quasi-power approaches zero (from below) and no blow-up is observed over the  time interval considered. This is the first observation of a recurring theme: numerics suggest that solutions with quasi-power near zero do not blow-up and exist globally.
%%%%%%%%%%%%%%%%%%%%%%%%%%%%%%%%%%%%%%%%%%%%%%%%%%
%%%%%%%%%%%%%%%%%%%%%%%%%%%%%%%%%%%%%%%%%%%%%%%%%%%
\subsection{PT collapse dynamics}
%%%%%%%%%%%%%%%%%%%%%%%%%%%%%%%%%%%%%%%%%%%%%%%%%%%%%%%%%%%%%%%%%%%%%%%%%%%
In this section we consider the PT reduction, in which case the underlying NLS equation is given in (\ref{eq11}). One can instead consider  reflection in the transverse direction by using Eq.~(\ref{y_nonlocal_sym_a}), or fully two-dimensional PT symmetry, in which case  there is nonlocality in both the $x$ and $y$ directions (see Eq.~(\ref{y_nonlocal_sym_b})). Note that for all results below, we set $v = 0$.
Consider the PT symmetric equation (\ref{eq11}), which is nonlocal only in $x$. We examine the blow-up dynamics when the nonlocality of the equation is evident, namely we consider $x_0 \not=0$ in (\ref{gaussian}). The result is an asymmetry in the initial condition, where $q(x,y,t)$ is centered at $(x_0,0)$ and $r(x,y,t)$ is centered at $(-x_0,0)$ due to the parity symmetry $x \rightarrow - x$. The relevant quasi-variance and conserved quantities for Gaussian PT initial conditions are: \\\\
{\bf PT:}
%%%%%%%%%%%%%%%%%%%%%%%%%%%%%%%%%%%%%%%%%%%%%%%%%%%
\begin{align}
\label{V_PT}
&V(0)  =-\frac{\pi A^2 e^{-2 B x_0^2 } (B + C + 4 BC y_0^2)}{ 8 B^{3/2} C^{3/2}}, \\
\label{g1_PT}
&P  = - \frac{\pi A^2}{2 B^{1/2} C^{1/2}} e^{- 2  B x_0^2  } , \\
\label{g3_PT}
&H  =   \frac{\pi A^2  e^{- 4 B x_0^2 } \left[ A^2 + 2 e^{2 B x_0^2} \left( -C + B (4 B x_0^2 - 1) \right) \right] }{4 B^{1/2} C^{1/2}}  .
\end{align}
%%%%%%%%%%%%%%%%%%%%%%%%%%%%%%%%%%%%%%%%%%%%%%%%%%%
If the nonlocality is only in $y$ (see Eq.~(\ref{y_nonlocal_sym_a})), then the results are conceptually similar to those discussed below by swapping $x_0$ and $y_0$ 
as well as $C$ and $B$.
Notice that the quasi-power (\ref{g1_PT}) and quasi-energy (\ref{g3_PT}) are independent of $y_0$.
This is due to the fact that Eq.~(\ref{eq11}) is nonlocal in $x$, but local in $y$.
This fact will change when the full two-dimensional PT symmetry is imposed.
A key difference from the stationary RT (and LOC) conserved quantities in (\ref{g1_NLS})-(\ref{g3_NLS}), is that (\ref{g1_PT})-(\ref{g3_PT}) depend on the position of the Gaussian (recall  $v = 0, n = 0$).
Examining the quasi-power in (\ref{g1_PT}), one sees that it decreases as $|x_0|$ increases, unlike the RT/LOC case in (\ref{g1_NLS}). Physically, increasing the $|x_0|$ separates the peaks of $q $ and $r$. As a result, the overlap of their exponential tails diminishes. Again, notice that ${\rm sgn}(V(0)) = -1$.  The quasi-energy invariant in (\ref{g3_PT}) can be either positive or negative.  However, for each case considered below, $H > 0$ so there will a time when $V(t)= 0$.

\begin{figure}
     \centering
     \includegraphics[scale=0.5]{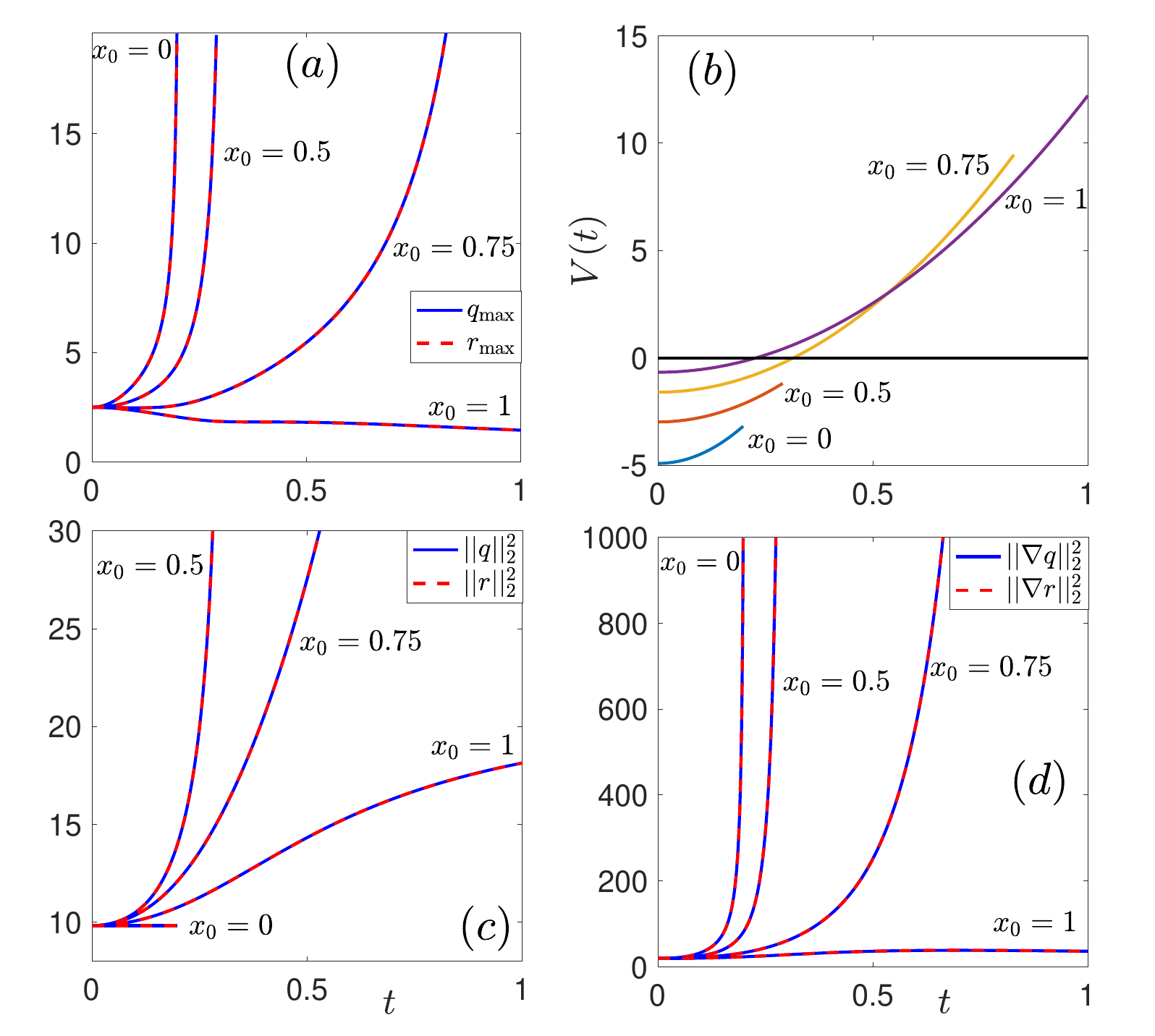}
\caption{Time evolution of the (a) maximum magnitude (\ref{eqqmax}) and (b) quasi-variance (\ref{eq53}) for the partial PT NLS equation (\ref{eq11}) and Gaussian initial condition (\ref{gaussian}) with different values of $x_0$. Also shown are the (c) $L^2$ norm (\ref{l2norm_define})  and (d) $L^2$ norm of the gradient (\ref{l2norm_grad_define}).  Blow-up is eventually avoided as $x_0$ increases. The other parameters  are $A = 2.5, B = C = 1, y_0 = 0, v = 0$. Note that $H >0 $ for each case considered here.}  
        \label{PT_collapse_summary}
\end{figure}

Some numerical findings are summarized in Fig.~\ref{PT_collapse_summary} for the NLS equation with PT symmetry in the $x$-direction only (see Eq.~(\ref{eq10})). When the centers of $q$ and $r$ are relatively close to each other ($0 < |x_0| \ll 1 , |P| \gg 1$), the solutions appear to collapse in finite time with their magnitudes (\ref{eqqmax}) rapidly growing. However, when the centers are sufficiently separated ($|x_0| \gg 1 , |P| \ll 1$), a  blow-up event is no longer observed. Instead, the solutions are simply observed to spread (diffract) and lose magnitude. Examining the quasi-variance in Fig.~\ref{PT_collapse_summary}(b), it is seen that $V(t)$ going to zero in finite time does not necessarily imply collapse in finite time. This is in contrast to the classical NLS equation, where the sign-definite nature of $V(t)$ can indicate  collapse.
%%%%%%%%%%%%%%%%%%%%%%%%%%%%%%%%%%%%%%%%%%%%%%%%%%
%%%%%%%%%%%%%%%%%%%%%%%%%%%%%%%%%%%%%%%%%%%%%%%%%%%
To shed more light on the dynamics, we monitor the time evolution of the $L^2$ norm for $q$ and $r$, defined by
%%%%%%%%%%%%%%
\begin{equation}
\label{l2norm_define}
|| f ||_2^2 = \iint_{\mathbb{R}^2} |f(x,y,t)|^2 ~ dx ~dy ,
\end{equation}
%%%%%%%%%%%%%%
as well as their gradient norm 
%%%%%%%%%%%%%%
\begin{equation}
\label{l2norm_grad_define}
|| \nabla f ||_2^2 = \iint_{\mathbb{R}^2} \left(  |  f_x |^2 + |  f_y |^2 \right) ~ dx ~dy ,
\end{equation}
%%%%%%%%%%%%%%
where $f_x \equiv \partial f / \partial x$ and $f_y \equiv \partial f / \partial y$. Traditionally, the gradient norm (\ref{l2norm_grad_define}) going to zero is used to indicate a sharply growing amplitude %the collapsing cross-section 
of a localized mode through an uncertainty principle \cite{Sulem,Fibich}. Note that, unlike the LOC case, 
the $L^2$ norm of $q$ and $r$ is, in general, {\it not} a conserved quantity. For such 
genuinely PT initial conditions ($x_0 \not=0$), the $L^2$ norm of the solution is not a constant of motion (see Fig.~\ref{PT_collapse_summary}(c)).
Figures~\ref{PT_collapse_summary}(c-d) show the time evolution of the $L^2$ norm of $q,r$ as well as their gradient norms. 
It is interesting to note that the non-conserved versions of these quantities blow-up at the same time as the maximal amplitudes shown in Fig.~\ref{PT_collapse_summary}.
% As such, we are going to monitor their evolution for the PT case.
 Moreover, the gradient norms do appear to grow unboundedly in the collapse cases (see Fig.~\ref{PT_collapse_summary}(d)). On the other hand, when the solution does not collapse (see Figs.~\ref{PT_collapse_summary}, %and \ref{L2_evolutions_resolved}, 
$x_0 = 1$), the gradient norm remains bounded for the time scales considered. The combination of peak magnitude  (Fig.~\ref{PT_collapse_summary}(a)) and  gradient (Fig.~\ref{PT_collapse_summary}(d)) going to infinity, indicate a collapse event is indeed taking place.
%%%%%%%%%%%%%%%%%%%%%%%%%%%%%%%%%%%%%%%%%%%%%%%%%%%
%\begin{figure}
%     \centering
%     \includegraphics[scale=0.6]{L2_evolutions_resolved.eps}
%\caption{The (a) $L_2$ norm (\ref{l2norm_define})  and (b) $L_2$ gradient norm (\ref{l2norm_grad_define}) of Gaussian PT solutions (\ref{gaussian}). All parameters  are the same as in Fig.~\ref{PT_collapse_summary}.}  
%        \label{L2_evolutions_resolved}
%\end{figure}
%%%%%%%%%%%%%%%%%%%%%%%%%%%%%%%%%%%%%%%%%%%%%%%%%%%%

%%%%%%%%%%%%%%%%%%%%%%%%%%%%%%%%%%%%%%%%%%%%%%%%%%
\begin{figure}
     \centering
     \includegraphics[scale=0.45]{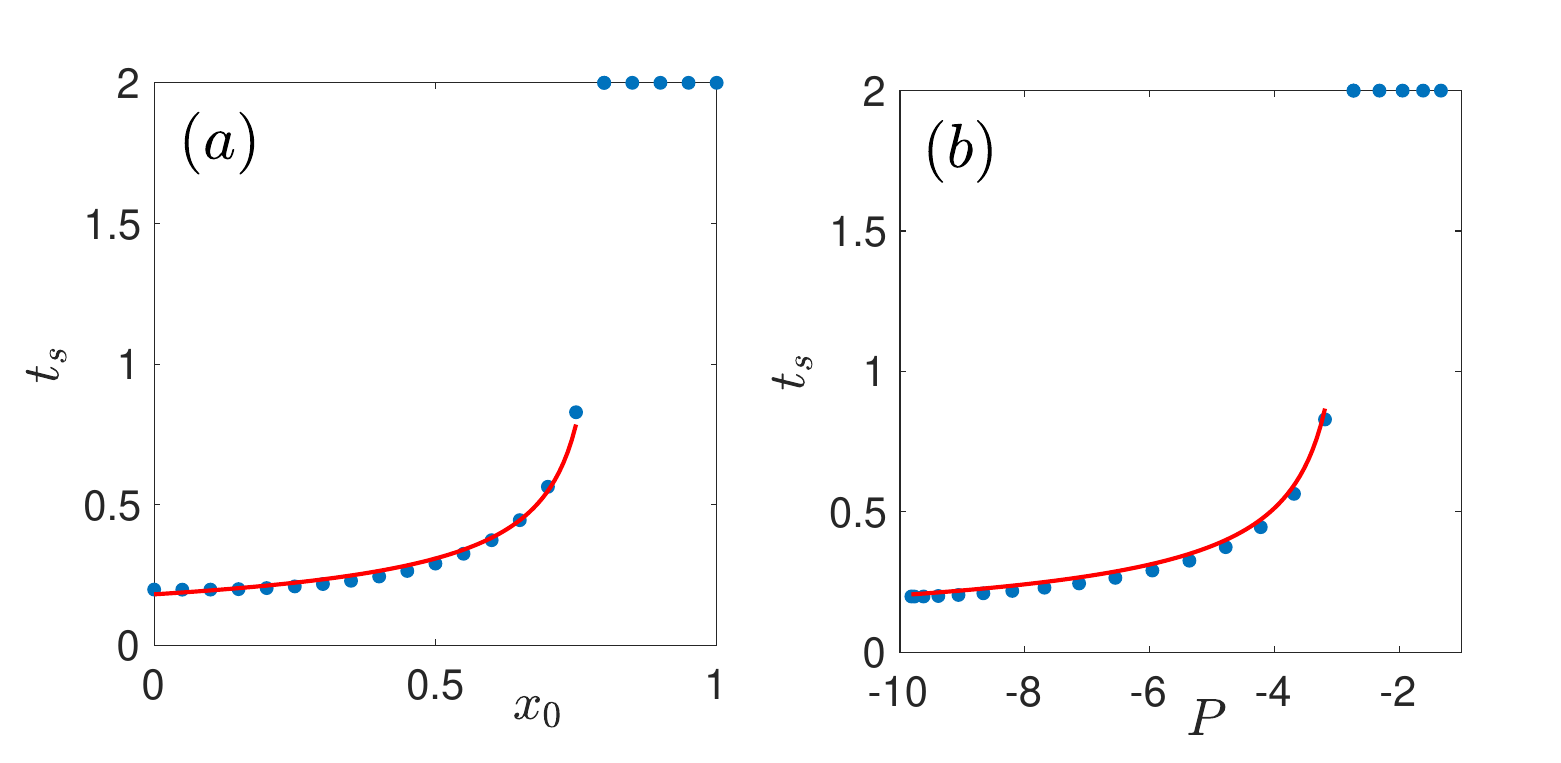}
\caption{The singularity time (\ref{singularity_time_define}) for partial PT Gaussian initial data (\ref{gaussian}) with $A = 2.5, B = C = 1, y_0 = 0$ and for different values of (a) $x_0$ and (b) $P$. Blue dots are numerical data. The red curves are given in (\ref{collapse_vert_asym}), respectively, with (a) $C_x = 0.18, K_x = -0.02, x_s = 0.8$ and (b) $C_P = 0.6, K_P = -0.02, P_s = -2.73$.} %The insets show a closer view near the singularity time. }  
        \label{Collapse_Time_vs_x0_1D}
\end{figure}
%%%%%%%%%%%%%%%%%%%%%%%%%%%%%%%%%%%%%%%%%%%%%%%%%%%

The results in Fig.~\ref{PT_collapse_summary} %and \ref{L2_evolutions_resolved} 
suggest that for fixed amplitude and width, changing the Gaussian centers amounts to controlling the collapse occurrence. This is due to the fact that the quasi-power directly depends on $x_0$. Hence, for fixed amplitude and width ($A,B,C$ fixed) that leads to collapse when $x_0  = 0$, moving the initial Gaussian center  can avoid collapse. What Fig.~\ref{Collapse_Time_vs_x0_1D} illustrates is the transition between collapsing solutions, corresponding to $x_0 \lesssim 0.8$, and those that do not collapse by $t = 2$ ($x_0 \gtrsim 0.8$). 
%%%%%%%%%%%%%%%%%%%%%%%%%%%%%%%%%%%%%%%%%%%%%%%%%%
\begin{figure}
     \centering
     \includegraphics[scale=0.6]{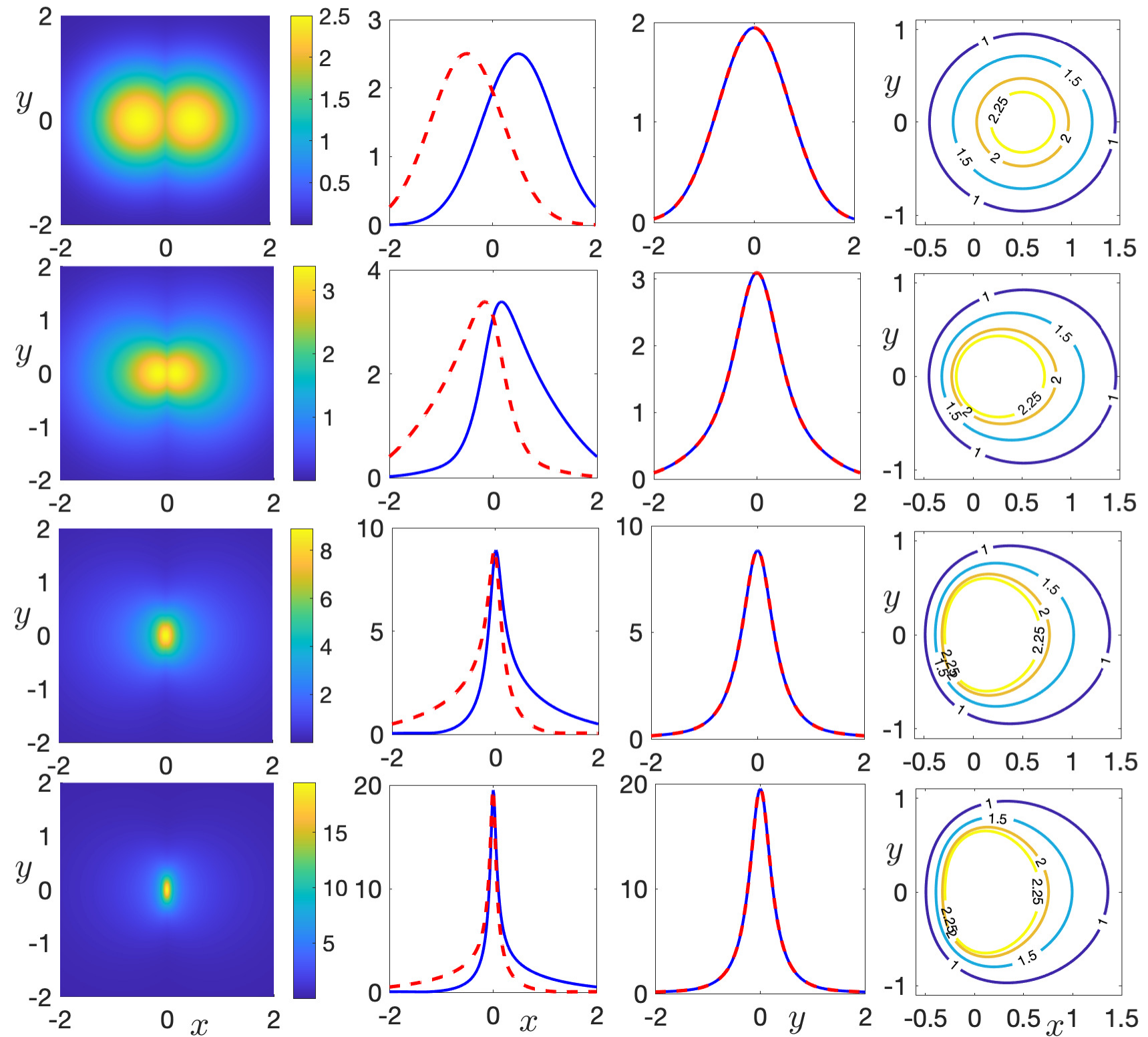}
\caption{Collapse profile snapshots for the partial (in $x$ direction) PT system shown in Fig.~\ref{Collapse_Time_vs_x0_1D} with $x_0 = 0.5$. Time evolves from top-to-bottom. Rows correspond to $t= 0, 0.15, 0.26, 0.29$, respectively. First (leftmost) column:  magnitudes $|q(x,y,t)|$ and $|r(x,y,t)|$. Second column: $y = 0$ profile cut, $|q(x,0,t)|$ (solid) and $|r(x,0,t)|$ (dashed). Third column: $x = 0$ profile slice, $|q(0,y,t)|$ (solid) and $|r(0,y,t)|$ (dashed). Fourth column: level curve contours denoting fixed magnitudes $|q(x,y,t)| = c$
with values of $c$: 1, 1.5, 2, 2.25. The level curves for $|r(x,y,t)| $ are similar, just reflected about $x = 0$. }  
        \label{profile_summary_1DPT}
\end{figure}
%%%%%%%%%%%%%%%%%%%%%%%%%%%%%%%%%%%%%%%%%%%%%%%%%%%
 The  solution appears to collapse  when $P \lesssim -2.7 $ (equivalently, $|P| \gtrsim 2.7$), and on these numerical time scales,  collapse is not detected for $P \gtrsim -2.7 $ (equivalently, $|P| \lesssim 2.7$). These simulations  seem to suggest there is a critical quasi-power condition   for establishing global existence \cite{Weinstein}. That is, it may be the case that for sufficiently small quasi-power, i.e.  $0 > P >  P_{\rm cr}$, global existence can be guaranteed.
Further scrutinizing the numerical data in Fig.~\ref{Collapse_Time_vs_x0_1D}, it appears there is a functional relationship between $t_s$ and $x_0$ or $P$. 
They are approximately given by
%%%%%%%%%%%%
\begin{equation}
\label{collapse_vert_asym}
t_s(x_0) \approx \frac{C_x}{(x_s - x_0)^{1/2}} + K_x , ~~ t_s(P) \approx \frac{C_P}{(P_s - P)^{1/2}} + K_P ,
\end{equation}
%%%%%%%%%%%%
where $x_0 < x_s , P < P_s $, and $C_x,C_P,K_x,K_P$ are fitting parameters. What motivates this class of functions is  
the presence of a vertical asymptote in both $x_0$ and $P$. Fitting parameters were chosen through trial and error to give a good approximation of the data.

With this in mind, a summary of the wavefunction profiles for the partial (in $x$) PT NLS equation is shown in Fig.~\ref{profile_summary_1DPT}. This particular evolution corresponds to the $x_0 = 0.5$ case depicted in Fig.~\ref{PT_collapse_summary}. We refer to this process  as {\it coalescence to collapse}. That is, the two initially separated peaks merge together before the cumulative state collapses (see left two columns). Furthermore, there does not appear to be any obvious self-similarity in the collapsing spatial profile. The discrepancy between the $y = 0$ and $x = 0$ profile cuts (middle two columns) highlights the absence of a radially dependent blow-up process. This is in contrast to the classical NLS shown in Fig.~\ref{profile_summary_NLS}. The rightmost column in Fig.~\ref{profile_summary_1DPT}  shows level curves of $|q(x,y,t)|$. Initially, the contours are radial about the center $(0.5,0)$. However, the contours lose their radial symmetry and become steeper on the left side (near the origin), and less steep on the right side (away from the origin). The functional shape for $|r(x,y,t)|$ can be obtained by reflection 
about $x = 0$, that is, $|r(x,y,t)| = |q(-x,y,t)|$. 

 %%%%%%%%%%%%%%%%%%%%%%%%%%%%%%%%%%%%%%%%%%%%%%%%%%
\begin{figure}
     \centering
     \includegraphics[scale=0.4]{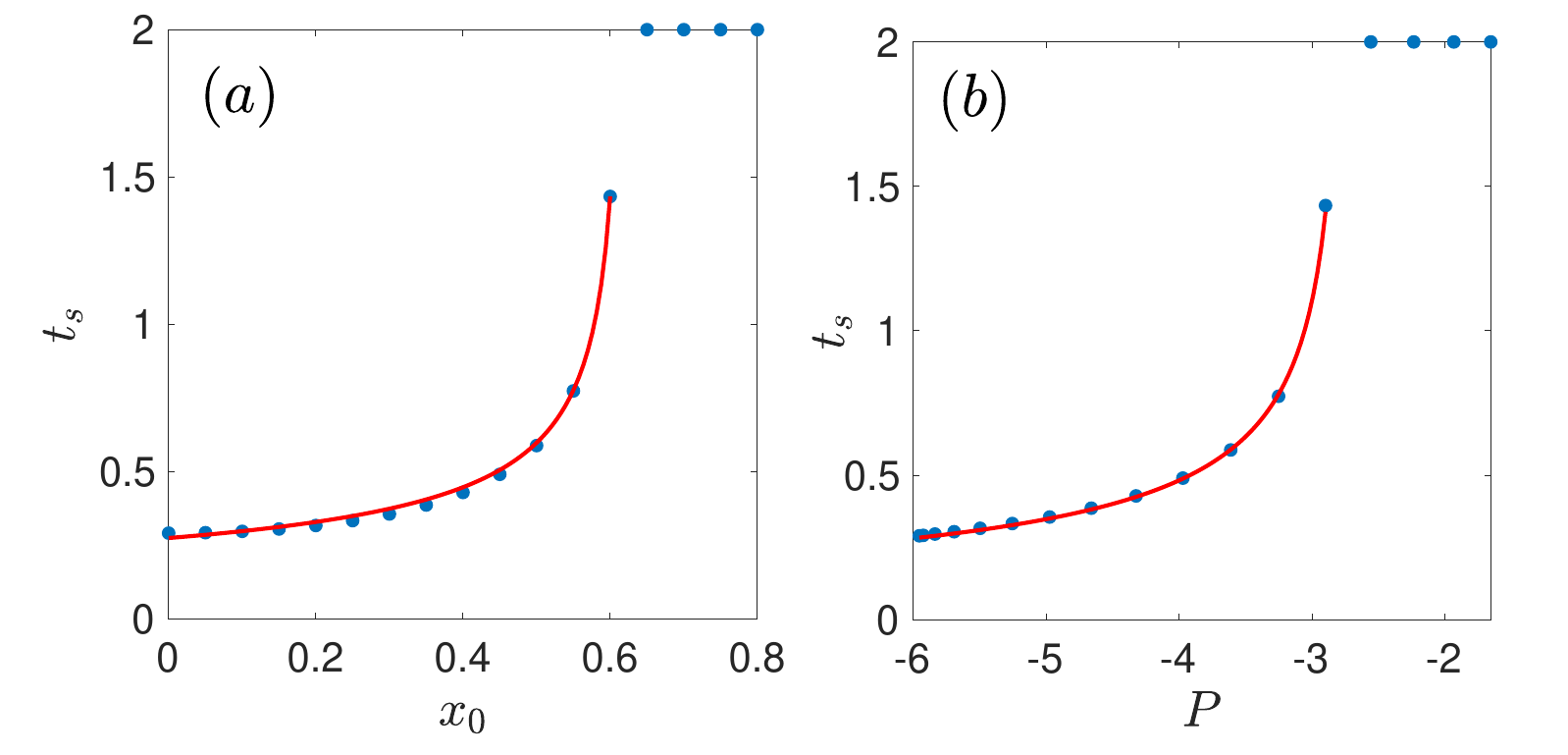}
\caption{The singularity time (\ref{singularity_time_define}) for Gaussian initial data (\ref{gaussian}) 
in full (reflection in both $x$ and $y$ directions) PT system with $A = 2.5, B = C = 1, y_0 = 0.5$, for different values of (a) $x_0$ and (b) $P$. Blue dots are numerical data. The red curves are given in (\ref{collapse_vert_asym}), respectively, with (a) $C_x = 0.18, K_x = -0.02, x_s = 0.62$ and (b) $C_P = 0.6, K_P = -0.05, P_s = -2.73$. The insets show a closer view near the singularity time. }  
        \label{Collapse_Time_vs_x0_2D}
\end{figure}
%%%%%%%%%%%%%%%%%%%%%%%%%%%%%%%%%%%%%%%%%%%%%%%%%%%

%\jtc{While the previous discussion suggests an absence of self-similarity in the transverse profile, the results in Fig.~\ref{} indicate an overall amplitude growth as through the collapse time is a simple pole. In Fig.~\ref{} we place fits on the maximum magnitudes, originally shown in Fig.~\ref{PT_collapse_summary}(a). In each case shown, a fit of the form 
%%%%%%%%%%%%%%%%%%%
%\begin{equation}
%L(t) = \frac{C}{(t_c - t)} , ~~ t < t_c
%\end{equation}
%%%%%%%%%%%%%%%%%%%
%is applied for some constant $C$. Near the collapse time, these fits agree well. Indeed, the NLS case ($x_0 = 0$) is known to collapse this way, but it is noteworthy that the PT collapse cases also seem to follow this trend. 
%}

Finally, we discuss the collapse dynamics for the fully (reflection in both $x$ and $y$ directions) PT-NLS equation 
%%%%%%%%%%%%%%%%%%
\begin{equation}
    iq_t(x,y,t)=\Delta q(x,y,t) + 2 q^2(x,y,t) q^\star(-x,-y,t) ,
\end{equation}
%%%%%%%%%%%%%%%%%%
where $q$ and $r$ are related by (\ref{y_nonlocal_sym_b}). The overall picture of blow-up in finite time persists for the full PT symmetric case as well. Indeed, Fig.~\ref{Collapse_Time_vs_x0_2D} summarizes the  apparent collapse dynamics when $y_0 = 1/2$. Notably, the empirically obtained critical power from the 
partial (reflection in the $x$ direction only) PT symmetric case $(P_{\rm cr} \approx -2.73)$ is also a good approximation here. This is  more evidence that there  exists a critical power, $P_{\rm cr} $, such that for $0 > P > P_{\rm cr} $  there is global existence.

\begin{figure}
     \centering
     \includegraphics[scale=0.6]{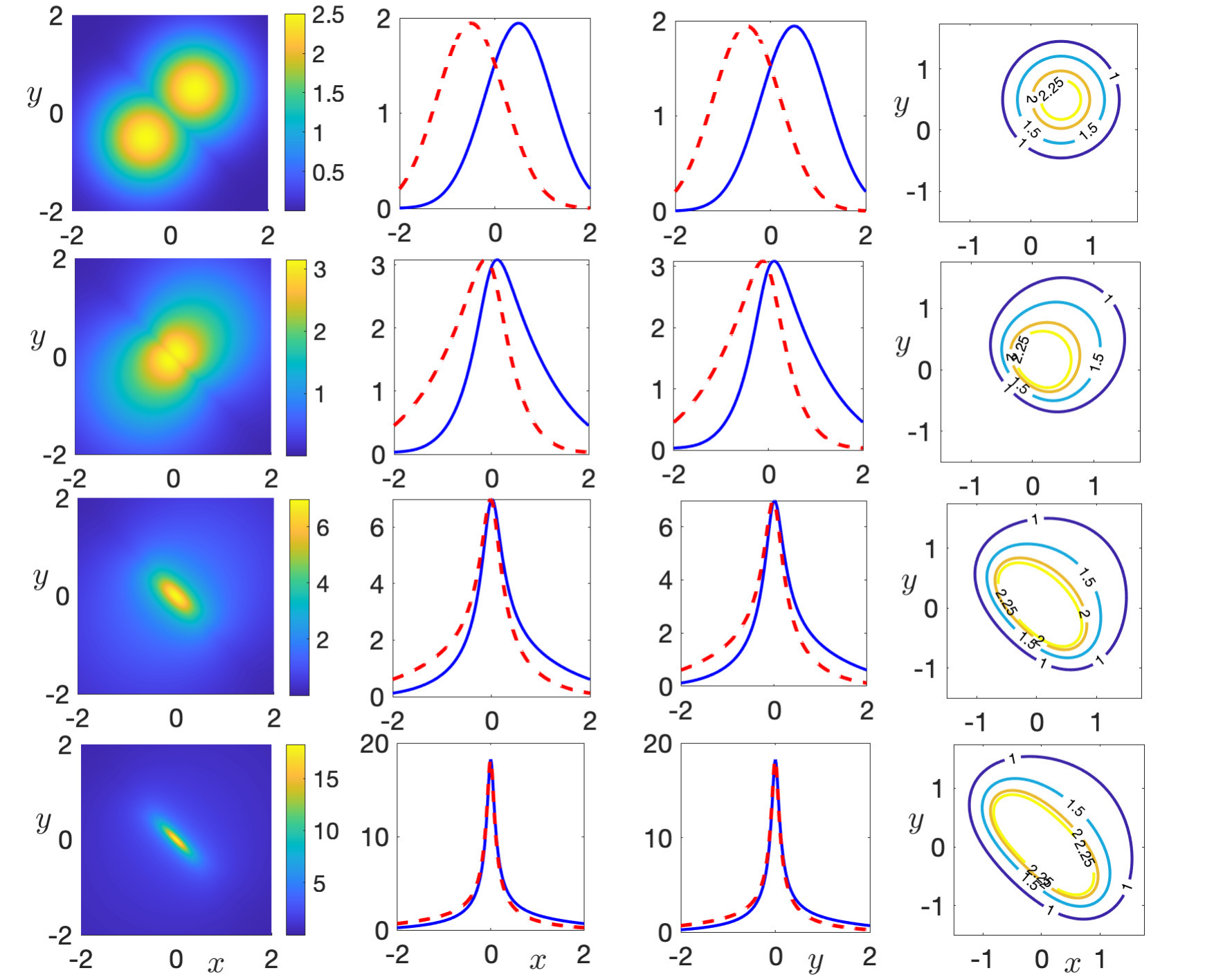}
\caption{Collapse profile snapshots for the full PT system (\ref{full_PT_r}) shown in Fig.~\ref{Collapse_Time_vs_x0_2D} with $(x_0,y_0) =( 0.5,0.5)$. Time evolves from top-to-bottom. Rows correspond to $t= 0, 0.25, 0.45, 0.58$, respectively. First (leftmost) column:  magnitudes $|q(x,y,t)|$ and $|r(x,y,t)|$. Second column: $y = 0$ profile cut, $|q(x,0,t)|$ (solid) and $|r(x,0,t)|$ (dashed). Third column: 
$x = 0$ profile cut, $|q(0,y,t)|$ (solid) and $|r(0,y,t)|$ (dashed). Fourth column: level curve contours denoting fixed magnitudes $|q(x,y,t)| = c$ with values of $c$: 1, 1.5, 2, 2.25. The level curves for $|r(x,y,t)| $ are similar, just reflected about $y =- x$.}  
        \label{profile_summary_2DPT}
\end{figure}

The spatial profile of a typical collapsing full PT wavefunction is shown in Fig.~\ref{profile_summary_2DPT}. Again, a coalescence to collapse process is observed. The two peaks fuse together on their way to an apparent collapse of the total state. We observe an unexpected symmetry in the $y = 0$ and $x = 0$ profile cuts (middle columns), where the two profiles look nearly identical. The level curves in the rightmost column again reveal a loss of the radially-symmetric initial profile.

%%%%%%%%%%%%%%%%%%%%%%%%%%%%%%%%%%%%%%%%%%%%%%%%%%
%%%%%%%%%%%%%%%%%%%%%%%%%%%%%%%%%%%%%%%%%%%%%%%%%%%
%%%%%%%%%%%%%%%%%%%%%%%%%%%%%%%%%%%%%%%%%%%%%%%%%%%%%%%%%%%%%%%%%
\subsection{General $q,r$ system}
\begin{figure}
     \centering
     \includegraphics[scale=0.55]{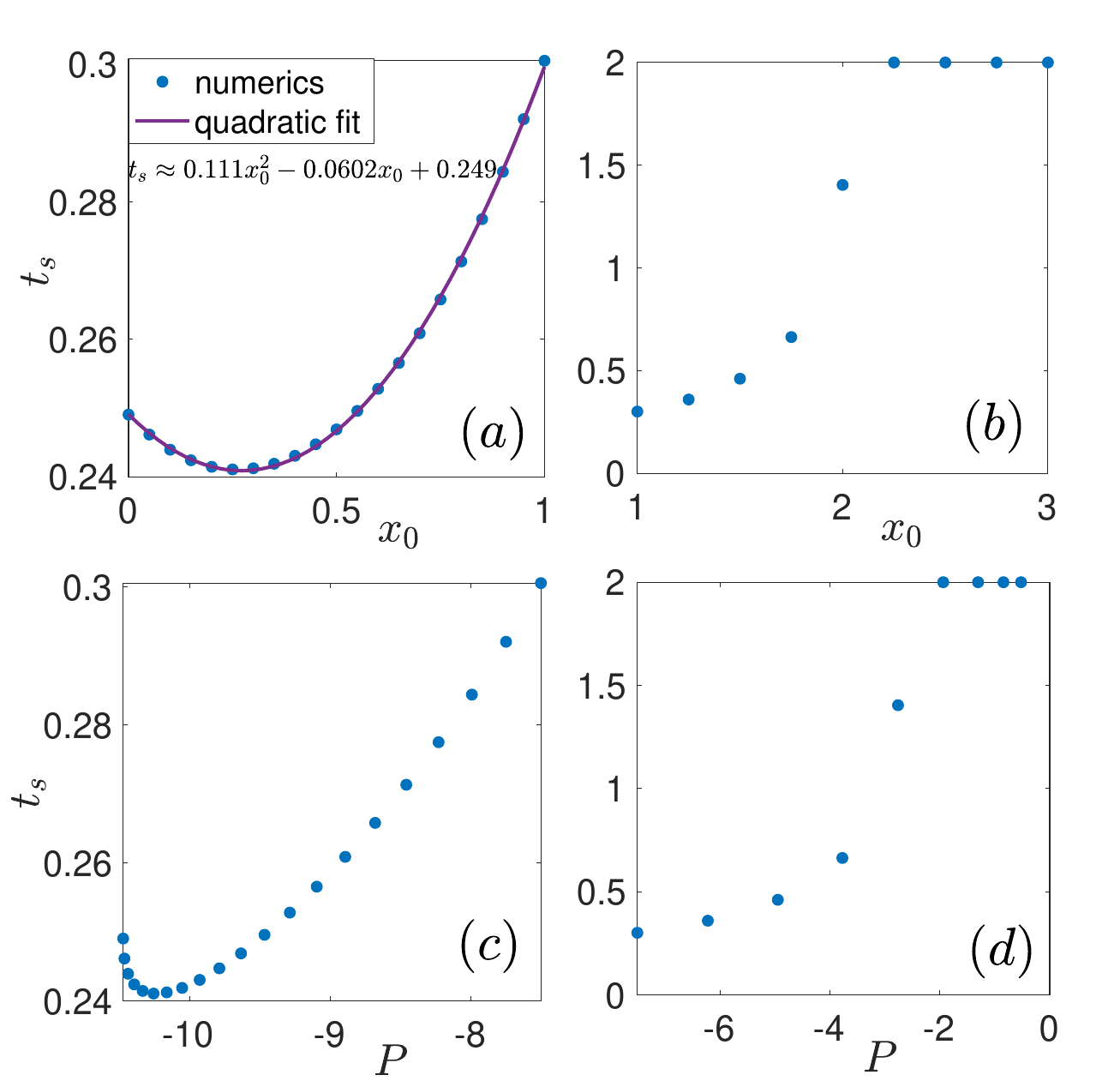}
\caption{The singularity time (\ref{singularity_time_define}) for the general $q,r$ system with initial data (\ref{general_qr_IC}), for different values of (a,b) $x_0$ and (c,d) $P$. Blue dots are indicate numerical data. A least squares quadratic fit is applied in panel (a).}  
        \label{general_qr_summary}
\end{figure}

%%%%%%%%%%%%%%%%%%%%%%%%%%%%%%%%%%%%%%%%%%%%%%%%%%%%%%%%%%%%%%%%%%
Motivated by the unexpected results found in the previous sections, here, we briefly study singularity formation for the general $q,r$ system in the absence of any local or nonlocal reductions between the $q(x,y,t)$ and $r(x,y,t)$ functions.
% The one-dimensional $q,r$ system (\ref{eq1})-(\ref{eq2}) supports many integrable reductions. We pose the question, why not study collapse in the  
%generic 2+1 $q,r$ system in its own right?
To do so, we simulate system (\ref{eq1})-(\ref{eq2}) subject to the following Gaussian initial conditions:
%%%%%%
\begin{equation}
\label{general_qr_IC}
q(x,y,0)  = A_1 e^{- B_1(x-x_0)^2 - C_1 (y - y_0)^2} , ~~~~~~ r(x,y,0)  =  -A_2 e^{- B_2x^2 - C_2 (y - y_0)^2} ,
\end{equation}
%%%%%
where $A_1 = 2.5, B_1 = C_1 = 1, y_0 = 0.5$ and $A_2 = 2, B_2 = C_2 = 0.5.$ There is, intentionally, no obvious symmetry reduction between these two functions. 
 The resulting blow-up dynamics are summarized in Fig.~\ref{general_qr_summary}. Similar to previous cases, initial data with very negative quasi-power, $P$, tend to blow-up, whereas, at small quasi-power the solutions do not appear to blow-up. %For $x_0 \ge 2.25$ ($P \ge -1.937$), which appear to not blow-up. %, we also ran the simulations up to time $t = 5$ \zhm{CHECK THE NUMBERS} and still did not observe any collapse-like dynamics. 
An intriguing feature observed here is the presence of a critical quasi-power ($x_0 \approx 0.25, P \approx -10.26$) at which the solution blows-up in the shortest amount of time. Compared to the previous cases, this is unusual since the collapse trend does not follow a monotonic trajectory.  
Furthermore, in Fig.~\ref{general_qr_summary}(a) the shape resembles that of a quadratic function. Motivated by this, we applied a least squares fit and obtained good agreement between the parabola and  numerical data. 
 %%%%%%%%%%%%%%%%%%%%%%%%%%%%%%%%%%%%%%%%%%%%%%%%%%
%%%%%%%%%%%%%%%%%%%%%%%%%%%%%%%%%%%%%%%%%%%%%%%%%%%
\section{Transverse instability analysis}
\label{linearstab}
%%%%%%%%%%%%%%%%%%%%%%%%%%%%%%%%%%%%%%%%%%%%%%%%%%%%
In this section, we study the linear stability of line soliton solutions when subject to transverse perturbations that depend on both $x$ and $y$ coordinates. This approach allows us to look at the issue of collapse from a rather different angle, i.e., throught transverse instability.
To begin, consider weakly perturbed solutions of Eqs.~(\ref{eq1}) and (\ref{eq2}) in the form
%%%%%%%%%%%%%%%%%%%%%%%%%%%%%%%%%%%%%%%%%%%%%%%%%%%%
\begin{align} \label{eq26}
q(x,y,t) & =[q_0(x)+\varepsilon Q(x,y,t)]e^{-i\mu t} , \\
 \label{eq27}
r(x,y,t) &=[r_0(x)+\varepsilon R (x,y,t)]e^{i\mu t} ,
\end{align}
%%%%%%%%%%%%%%%%%%%%%%%%%%%%%%%%%%%%%%%%%%%%%%%%%%%%
where $0 <\varepsilon \ll 1$ is a dimensionless parameter used to control the magnitude of the perturbation and $Q, R$ are  transverse perturbation that depend on $x, t$ {\it and} $y$. 
Next, we substitute these perturbations into the governing equations (\ref{eq1})-(\ref{eq2}) and linearize about the line soliton solution.  
That is, neglect all terms that are $O(\varepsilon^2)$. In the $q,r$ framework, the derivation is same for all  three reductions. Thus, we find 
%%%%%%%%%%%%%%%%%%%%%%%%%%%%%%%%%%%%%%%%%%%%%%%%%%%%
\begin{align} \label{eq28}
 iQ_{t} & = \Delta Q - ( \mu+4q_0r_0) Q - 2q_0^2R, \\
 \label{eq29}
- iR_{t} & =  \Delta R - (\mu+4q_0r_0) R - 2r_0^2Q  .
\end{align}
%%%%%%%%%%%%%%%%%%%%%%%%%%%%%%%%%%%%%%%%%%%%%%%%%%%%
In Eqs.~(\ref{eq28})-(\ref{eq29}) the solution $q_0$ is given by (\ref{eq16})-(\ref{eq18}); note that $r_0(x) = - q_0^\star(-x)$ in each case.
In order to effectively eliminate the $y$ and $t$ dependence, the functional forms of the $Q$ and $R$ functions must be  chosen so that the symmetries (\ref{eq6}), (\ref{eq8}), or (\ref{eq10}) are satisfied.
Below are the partial Fourier transform representations we use, which respect the associated symmetries: 
\\\\ {\bf LOC}
%%%%%%%%%%%%%%%%%%%%%%%%%%%%%%%%%%%%%%%%%%%%%%%%%%%%
\begin{align} 
\label{eq30}
&Q(x,y,t) =
\int \left( f(x,l)e^{i[ ly+\omega(l)t]}+g^{\star}(x,l) e^{-i[ ly+\omega^{\star}(l)t]} \right) dl
%%%%%%%%%%%%%%%%%%%%%%%%%%%%%%%%%%%%%%%%%%%%%%%%%%%%
 \\ \nonumber
&R (x,y,t)  = - Q^\star (x,y,t) ;
\end{align}
%%%%%%%%%%%%%%%%%%%%%%%%%%%%%%%%%%%%%%%%%%%%%%%%%%%%
{\bf RT}
%%%%%%%%%%%%%%%%%%%%%%%%%%%%%%%%%%%%%%%%%%%%%%%%%%%%
\begin{align} 
\label{eq30a}
&Q(x,y,t)  =  \int  \left(  f(x,l)e^{i[ ly+\omega(l)t]}+g(x,l)e^{i[ ly- \omega(l)t]} \right) dl 
 % \label{eq31a}
 \\ \nonumber
&R(x,y,t) = -Q(x,y,-t) ;
\end{align}
%%%%%%%%%%%%%%%%%%%%%%%%%%%%%%%%%%%%%%%%%%%%%%%%%%%%%
{\bf PT}
%%%%%%%%%%%%%%%%%%%%%%%%%%%%%%%%%%%%%%%%%%%%%%%%%%%%
\begin{align} 
\label{eq30b}
&Q(x,y,t)  = 
Q_{{\bf LOC}},  
\\  \nonumber
&R(x,y,t) = - Q^\star(-x,y,t) ,
\end{align}
%%%%%%%%%%%%%%%%%%%%%%%%%%%%%%%%%%%%%%%%%%%%%%%%%%%%
where $Q_{{\bf LOC}}$ is given in (\ref{eq30}).
In each case, $f(x,l)$ and $g(x,l)$ are assumed to be localized in $x$ and $l$, with transverse wavenumber $l$ and (complex) dispersion relation $\omega(l)$. 
Note that the nonlocal $y$ variants in Eqs.~(\ref{y_nonlocal_sym_a})-(\ref{y_nonlocal_sym_b}) will have a similar form, only with a $y \rightarrow - y$ replacement in the $R$ function.
Substituting the above forms into Eqs.~(\ref{eq28})-(\ref{eq29}), we are able to reduce the problem down from a partial differential equation to the boundary value eigenvalue problem
%%%%%%%%%%%%%%%%%%%%%%%%%%%%%%%%%%%%%%%%%%%%%%%%%%%%%%%
%%%%%%%%%%%%%%%%%%%%%%%%%%%%%%%%%%%%%%%%%%%%%%%%%%%%%%%
\begin{equation} \label{eq33}
\begin{pmatrix} 
0& L_1+l^2 \\
L_2+l^2&0
\end{pmatrix}
\quad
\begin{pmatrix} 
F \\
G
\end{pmatrix}= \omega \begin{pmatrix} 
F\\
G
\end{pmatrix} ,
\end{equation} 
%%%%%%%%%%%%%%%%%%%%%%%%%%%%%%%%%%%%%%%%%%%%%%%%%%%%%%%
where $L_1=\mu- 6 q_0^2(x)-\partial_x^2$ and $L_2=\mu- 2 q_0^2(x)-\partial_x^2$. Note that we have used the symmetry relation between $r_0$ and $q_{0}$.
%%%%%%%%%%%%%%%%%%%%%%%%%%%%%%%%%%%%%%%%%%%%%%%%%%%%%%%%%%%
%%%%%%%%%%%%%%%%%%%%%%%%%%%%%%%%%%%%%%%%%%%%%%%%%%%%%%%% 
The eigenfunctions $F,G$ in system (\ref{eq33}) are found by taking linear combinations of the functions $f,g$, namely
%%%%%%%%%%%%%%%%%%%%%%%%%%%%%%%%%%%%%%%%%%%%%%%%%%%%%%%
%%%%%%%%%%%%%%%%%%%%%%%%%%%%%%%%%%%%%%%%%%%%%%%%%%%%%%%
\begin{align*}
{\rm \bf{RT}:} ~~~ & F(x,l) =f(x,l)-  g(x,l) , \\
~~~ & G(x,l) =f(x,l)+  g(x,l). 
\end{align*}
%%%%%%%%%%%%%%%%%%%%%%%%%%%%%%%%%%%%%%%%%%%%%%%%%%%%%%%
\begin{align*}
{\rm \bf{PT}:} ~~~ & F(x,l)=f(x,l)-  g(-x,l) , \\  ~~~ & G(x,l)=f(x,l)+  g(-x,l).
\end{align*}
%%%%%%%%%%%%%%%%%%%%%%%%%%%%%%%%%%%%%%%%%%%%%%%%%%%%%%%
The eigenfunctions $F,G$ corresponding to the local (LOC) case  are %given by 
the same form %expression 
as the RT one.
If one instead considers nonlocality in $y$, i.e. (\ref{y_nonlocal_sym_a})-(\ref{y_nonlocal_sym_b}), then the is no change to system (\ref{eq33}) due to the second-order  derivative in $y$; that is, $\partial_y^2 \rightarrow - l^2$ regardless of nonlocality in $y$.

Observe for the exponential forms given in Eqs.~(\ref{eq30})-(\ref{eq30b}) that, for a fixed $l$, the eigenvalues $\omega (l)$ of (\ref{eq33}) with nonzero imaginary part will exhibit exponential growth or decay in $t$. Furthermore, note that if $\omega$ is an eigenvalue of Eq.~(\ref{eq33}), then so is $\omega^{\star}$ since the system is invariant under the symmetry: ${\rm c.c.} + x \rightarrow - x$. As a result, if there exists an unstable eigenvalue, i.e. ${\rm Im} \left\{\omega \right\} \not=0$, then there is a  pair: one growing eigenmode and one decaying; the growing mode concerns us.
We refer to  eigenvalue/eigenvector pairs with ${\rm Im} \left\{\omega \right\} \not=0$ as linearly unstable to transverse modulations with wavenumber $l$. Linearly unstable perturbations will grow exponentially until large amplitude nonlinear effects become significant. Regardless, we expect any linearly unstable mode to have a noticeable effect on the evolution of the fully nonlinear dynamics.
%%%%%%%%%%%%%%%%%%%%%%%%%%%%%%%%%%%%%%%%%%%%%%%%%%%%%%%
\begin{figure}
	\centering
	\includegraphics[height=0.5\textwidth]{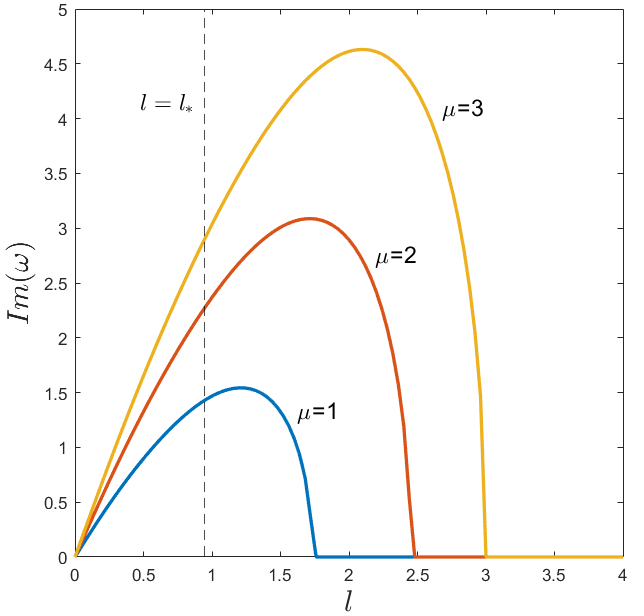}
	\caption{Transverse instability spectrum found by numerically solving Eq. (\ref{eq33}). All three solutions, as given by Eqs. (\ref{eq16})-(\ref{eq18}), have identical instability profiles. The value of $l_* =3 \pi / 10 \approx 0.94248$ is marked.}
	\label{f1}
\end{figure}
%%%%%%%%%%%%%%%%%%%%%%%%%%%%%%%%%%%%%%%%%%%%%%%%%%%%%%%

To determine linear stability, we numerically solve eigenvalue problem (\ref{eq33})  as a function of the transverse wavenumber $l$. This is accomplished through spectral collocation method where the the second derivative $\partial_x^2$ is approximated by spectral differentiation matrices on a large computational domain \cite{Trefethen}. The stability is computed for the solutions (\ref{eq16}), (\ref{eq17}), and (\ref{eq18}).
%%%%%%%%%%%%%%%%%%%%%%%%%%%%%%%%%%%%%%%%%%%%%%%%%%%%%%%
\begin{figure}
	\centering
	\includegraphics[scale = 0.55]{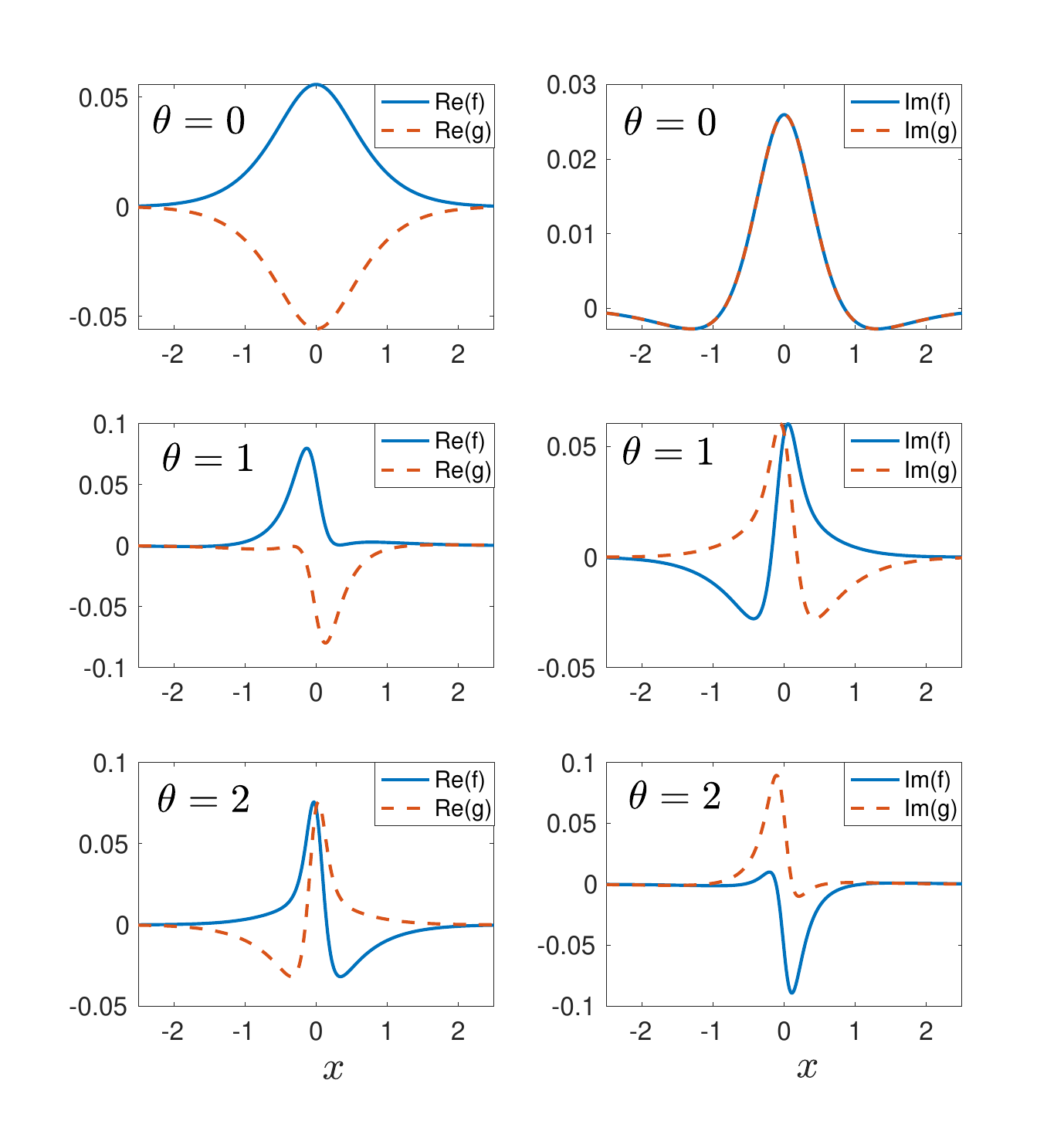}
	\caption{Eigenfunctions $f(x,l_*)$ and $g(x,l_*)$ obtained by solving eigenvalue problem (\ref{eq33}) for PT symmetric solutions (\ref{eq18}) at  $l_* =3 \pi / 10 \approx 0.94248$ and $\mu = 3$.  Note: these are formed from $F(x,l_*)$ and G$(x,l_*)$. The left (right) column shows the real (imaginary) part for different values of $\theta$. These eigenfunctions are used to produce the simulations in Fig.~\ref{evolution_snapshots}.}
	\label{f_eigs}
\end{figure}
%%%%%%%%%%%%%%%%%%%%%%%%%%%%%%%%%%%%%%%%%%%%%%%%%%%
The results of the stability analysis are shown in Fig.~\ref{f1}.  Our main finding is that the soliton  solutions of the nonlocal equations have the same instability spectra as the classical NLS solution. For that reason, we only show one instability curve for each soliton propagation constant $\mu$. Notably, each solution is unstable to slow transverse modulations, but is not expected to be unstable to rapidly varying transverse modulations.
As a secondary comment, observe that at $l = 0$, there is no linear instability expected. This case corresponds to a purely longitudinal ($x$-direction only) perturbation. To excite linear unstable modes, there needs to be a transverse dependence in the perturbation function. 
The eigenfunctions corresponding to a linearly unstable eigenvalue are shown in Fig.~\ref{f_eigs}. In the PT nonlocal case ($\theta\not = n\pi,  \pi \left( \frac{1}{2} + n \right) , n \in\mathbb{Z} $), the eigenfunctions take on an asymmetric %rather unexpected 
form. Specifically, we observe the eigenfunctions are apparently related via  $f(x,l) = \pm  g^{\star}(-x,l)$, where the sign  depends on $\theta$. Substituting this relationship into the full two-dimensional functions in (\ref{eq30b}), shows the PT symmetry $R(x,y,t) = - Q^{\star}(-x,y,t)$ is indeed satisfied.
%%%%%%%%%%%%%%%%%%%%%%%%%%%%%%%%%%%%%%%%%%%%%%%%%%%%%%%
%%%%%%%%%%%%%%%%%%%%%%%%%%%%%%%%%%%%%%%%%%%%%%%%%%%%%%%
%%%%%%%%%%%%%%%%%%%%%%%%%%%%%%%%%%%%%%%%%%%%%%%%%%%%%%%
\section{Slow modulation analysis}
\label{asymptotic}
%%%%%%%%%%%%%%%%%%%%%%%%%%%%%%%%%%%%%%%%%%%%%%%%%%%%
To support the above numerical findings and to better understand the transverse instability,  we develop, in this section, a perturbation theory valid in the long wavelength limit, i.e. $l \rightarrow 0$. The advantage of this approach is that we are able to derive analytic formulas for the perturbation eigenvalues $\omega(l)$. For the analysis below, we shall make a frequent use of the
inner product 
%%%%%%%%%%%%%%%%%%%%%%%%%%%%%%%%%%%%%%%%%%%%%%%%%%%%%%%
\begin{equation} \label{eq34a}
\langle \varphi_1,\varphi_2 \rangle = \int_{-\infty}^\infty \varphi_1(x) \varphi_2(x) dx  .
\end{equation}
%%%%%%%%%%%%%%%%%%%%%%%%%%%%%%%%%%%%%%%%%%%%%%%%%%%%%%%
Next, we expand the perturbation eigenfunctions $F,G$ as well as the eigenvalues $\omega(l)$ in an asymptotic series for small $l$, i.e. $0 < l \ll  1$:
%%%%%%%%%%%%%%%%%%%%%%%%%%%%%%%%%%%%%%%%%%%%%%%%%%%%%%%
\begin{equation} \label{eq36}
\omega=l\omega_1+ l^2\omega_2 + \dots
\end{equation}
%%%%%%%%%%%%%%%%%%%%%%%%%%%%%%%%%%%%%%%%%%%%%%%%%%%%%%%
\begin{equation} \label{eq37}
F(x,l)=F_0(x)+lF_1(x)+ l^2F_2(x)+ \dots
\end{equation}
%%%%%%%%%%%%%%%%%%%%%%%%%%%%%%%%%%%%%%%%%%%%%%%%%%%%%%%
\begin{equation} \label{eq38}
G(x,l)=G_0(x)+lG_1(x)+ l^2G_2(x)+ \dots ,
\end{equation}
%%%%%%%%%%%%%%%%%%%%%%%%%%%%%%%%%%%%%%%%%%%%%%%%%%%%%%%
where all the eigenfunctions $F_n$ and $G_n$ are assumed to be smooth. Collecting terms at the first three orders in  
$l$, we get: 
%%%%%%%%%%%%%%%%%%%%%%%%%%%%%%%%%%%%%%%%%%%%%%%%%%%%%%%
\begin{equation} \label{eq40}
O(1): \hspace{2mm} L_1 G_0=0, \hspace{1cm} L_2 F_0=0
\end{equation}
%%%%%%%%%%%%%%%%%%%%%%%%%%%%%%%%%%%%%%%%%%%%%%%%%%%%%%%
\begin{equation} \label{eq41}
O(l): \hspace{2mm} L_1 G_1=\mathcal{S}_1, \hspace{1cm} L_2 F_1=\mathcal{T}_1
\end{equation}
%%%%%%%%%%%%%%%%%%%%%%%%%%%%%%%%%%%%%%%%%%%%%%%%%%%%%%%
\begin{equation} \label{eq42}
O(l^2): \hspace{2mm} L_1 G_2=\mathcal{S}_2, \hspace{1cm} L_2 F_2=\mathcal{T}_2
\end{equation}
%%%%%%%%%%%%%%%%%%%%%%%%%%%%%%%%%%%%%%%%%%%%%%%%%%%%%%%
where $$ \mathcal{S}_1 =\omega_1F_0, \hspace{2mm} \mathcal{T}_1=\omega_1G_0 , $$
and $$\mathcal{S}_2=\omega_1F_1+\omega_2F_0 - G_0, \hspace{2mm} \mathcal{T}_2 =\omega_1G_1+\omega_2G_0 - F_0.$$
Our aim is to solve Eqs. (\ref{eq40})-(\ref{eq42}) successively.  
The solutions of (\ref{eq40}) are readily obtained from Eq.~(\ref{eqstat}), since
%%%%%%%%%%%%%%%%%%%%%%%%%%%%%%%%%%%%%%%%%%%%%%%%%%%%%%%
\begin{equation} \label{eq44}
L_1 (\partial_x q_0)=0, \hspace{1cm} L_2q_0=0 ,
\end{equation}
%%%%%%%%%%%%%%%%%%%%%%%%%%%%%%%%%%%%%%%%%%%%%%%%%%%%%%%
where the former equation is obtained by differentiating Eq.~(\ref{eqstat}) with respect to $x$. As a result, we see that a set of solutions is given by
%%%%%%%%%%%%%%%%%%%%%%%%%%%%%%%%%%%%%%%%%%%%%%%%%%%%%%%
\begin{equation} \label{eqeigfunc}
G_0 = c_1 q_{0x}(x) ~~~~~ F_0 = c_2 q_0(x) ,
\end{equation}
%%%%%%%%%%%%%%%%%%%%%%%%%%%%%%%%%%%%%%%%%%%%%%%%%%%%%%%
for arbitrary constant coefficients $c_1,c_2$. Differentiating Eq.~(\ref{eqstat}) with respect to the soliton eigenvalue $\mu$, we find $$G_1 = - c_2 \omega_1 \partial_{\mu} q_0.$$ 

%%%%%%%%%%%%%%%%%%%%%%%%%%%%%%%%%%%%%%%%%%%%%%%%%%%%%%%
\begin{figure}
	\centering
	\includegraphics[height=.3\textheight]{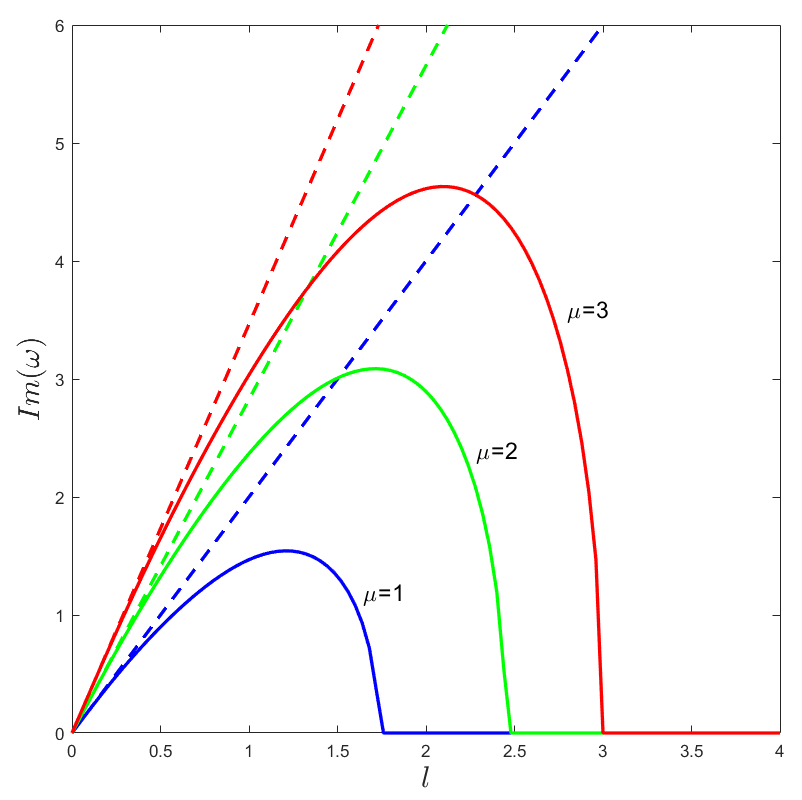}
	\caption{Comparison between numerical results obtained by solving Eq.~(\ref{eq33}) and the long wavelength approximation (\ref{eq49}) for different values of $\mu$. Notice that the results are independent of $\theta$ values in Eq.~(\ref{eq18}). All three solutions, as given by Eqs. (\ref{eq16})-(\ref{eq18}), have identical unstable spectral %instability 
profiles.}
	\label{f2}
\end{figure}
%%%%%%%%%%%%%%%%%%%%%%%%%%%%%%%%%%%%%%%%%%%%%%%%%%%%%%%
Next, we seek a solvability condition to determine the first correction to the perturbation eigenvalue in Eq.~(\ref{eq36}).
% To do so, we multiply Eqs.~(\ref{eq41}) and (\ref{eq42}) by $q_0(x)$. %in the LOC and RT cases or by $q^\star(-x)$ in the PT cases.
Multiplying the first (second) equation of (\ref{eq41}) by $G_{0}$ ($F_0$) yields no new information  about $\omega_1$ since
$$ \langle G_0 , L_1 G_1  \rangle =  0 = \langle G_0 , \mathcal{S}_1 \rangle ,$$
and
$$ \langle F_0 , L_2 F_1  \rangle =  0 = \langle F_0 , \mathcal{T}_1 \rangle ,$$
identically. Multiplying the second equation in (\ref{eq42}) by $F_0$, we observe that $\langle F_0 , L_2 F_2\rangle = \langle L_2 F_0 , F_2 \rangle = 0$ for the inner product in (\ref{eq34a}). As a result, we obtain the solvability condition 
%%%%%%%%%%%%%%%%%%%%%%%%%%%%%%%%%%%%%%%%%%%%%%%%%%%%%%%
\begin{equation} \label{eq48}
\langle F_0, \mathcal{T}_2 \rangle = \omega_1 \langle F_0, G_1 \rangle -  \langle F_0 ,  F_0  \rangle = 0 \; ,
\end{equation}
%%%%%%%%%%%%%%%%%%%%%%%%%%%%%%%%%%%%%%%%%%%%%%%%%%%%%%%
which  gives the following expression for the perturbation in the long wavelength limit
%%%%%%%%%%%%%%%%%%%%%%%%%%%%%%%%%%%%%%%%%%%%%%%%%%%%%%%
\begin{equation}\label{eq49}
\omega  = \pm i l \Omega_1  +O(l^2) , ~~~~~\Omega_1^2 =4\mu .
\end{equation}
%%%%%%%%%%%%%%%%%%%%%%%%%%%%%%%%%%%%%%%%%%%%%%%%%%%%%%%
Note that each solution in (\ref{eq16})-(\ref{eq18}) is found to give identical results here.
This result predicts that, in the long wavelength, the instability growth rate is proportional to $\mu^{1/2}$ %the soliton eigenvalue 
for each NLS equation and solution considered. Moreover, notice that this result does not depend on the parameter $\theta$ in solution (\ref{eq18}).
The numerical and analytical findings are compared in Fig. (\ref{f2}). There is excellent agreement in the small $l$ limit. 
 %%%%%%%%%%%%%%%%%%%%%%%%%%%%%%%%%%%%%%%%%%%%%%%%%%%%%%%%%%%%%
 %%%%%%%%%%%%%%%%%%%%%%%%%%%%%%%%%%%%%%%%%%%%%%%%%%
\section{Direct numerical simulations: Filamentation}
\label{cases}
%%%%%%%%%%%%%%%%%%%%%%%%%%%%%%%%%%%%%%%%%%%%%%%%%%
%%%%%%%%%%%%%%%%%%%%%%%%%%%%%%%%%%%%%%%%%%%%%%%%%%
In this section we perform several direct numerical simulations to confirm the linear stability results obtained in Secs. \ref{linearstab} and \ref{asymptotic}, as well as visualize  the development of  instability in the fully nonlinear system \cite{Klein2}. The analysis of the previous sections yields valid insight only when the perturbations $Q,R$ are not too large. In the linearly unstable cases, that assumption will most likely be violated given enough time and hence fully nonlinear simulations are needed.
%%%%%%%%%%%%%%%%%%%%%%%%%%%%%%%%%%%%%%%%%%%%%%%%%%
%%%%%%%%%%%%%%%%%%%%%%%%%%%%%%%%%%%%%%%%%%%%%%%%%%
%%%%%%%%%%%%%%%%%%%%%%%%%%%%%%%%%%%%%%%%%%%%%%%%%%
\begin{figure*}
     \centering
     \includegraphics[scale=0.6]{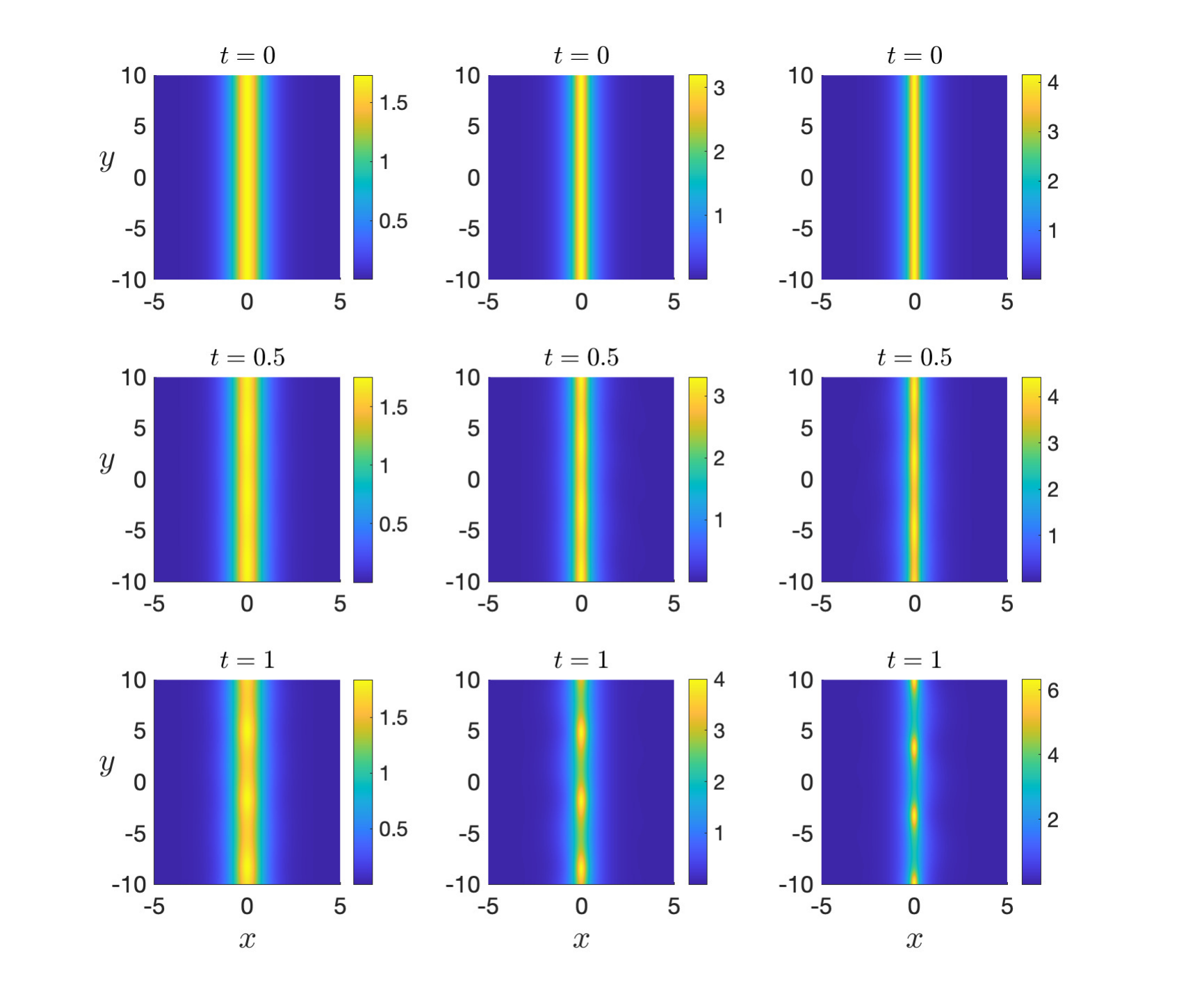}
        \caption{Transverse instability evolution of $|q(x,y,t)|$ for different values of $\theta$ and fixed $\mu = 3, \varepsilon = 0.1$. System (\ref{eq1})-(\ref{eq2}) is solved using initial conditions (\ref{eq26})-(\ref{eq27}) and PT system (\ref{eq30b}) for the eigenfunctions in Fig.~\ref{f_eigs}.  Columns,  left-to-right, correspond to $\theta=0,1, 2$, respectively. When $\theta = 0$, the evolutions  for all three NLS equations are  similar.} 
        \label{evolution_snapshots}
\end{figure*}
%%%%%%%%%%%%%%%%%%%%%%%%%%%%%%%%%%%%%%%%%%%%%%%%%%
%%%%%%%%%%%%%%%%%%%%%%%%%%%%%%%%%%%%%%%%%%%%%%%%%%
%%%%%%%%%%%%%%%%%%%%%%%%%%%%%%%%%%%%%%%%%%%%%%%%%%

A  fourth-order integrating Runge-Kutta factor method  with Fourier approximations of derivatives \cite{Yang} was implemented again to solve the $q,r$ system (\ref{eq1})-(\ref{eq2}). The computational window used is $ [-15 , 15] \times  [-10,10 ]$ for domain lengths 
$L_x = 30$ (with 2048 grid points) and $L_y= 20$ ((with 1400 grid points) while keeping the time-step $\delta t = 10^{-5}$.
Periodic boundary conditions are used in both direction to approximate the localized ($x$-direction) and periodic ($y$-direction) boundary behavior. 
%\jtc{The grid spacings $\delta x, \delta y$ are kept at the same values as the previous simulations.}
We find it considerably easier to integrate the $q,r$ system since the nonlocal system resembles a local one there. Contrast this with the scalar reductions (e.g.  Eq.~(\ref{eq11}))
 where one must account for nonlocality at each time step, e.g. $x \rightarrow -x$. The nonlocal nature of the systems is imposed through the initial conditions and preserved throughout the simulations. 
 %%%%%%%%%%%%%%%%%%%%%%%%%%%%%%%%%%%%%%%%%%%%%%%%%%%%%%%%%%%%%%%%%%%%%%%%%%%%%%
 
The initial conditions are taken from (\ref{eq26})-(\ref{eq27}) at $t = 0$. In the simulations below, we choose eigenfunctions corresponding to linearly unstable modes (see Figs.~\ref{f1} and \ref{f_eigs}). To do so, the eigenvalue problem in Eq.~(\ref{eq33}) is solved numerically and the eigenfunctions $F(x,l),G(x,l)$ corresponding to ${\rm Im}\left\{ \omega(l) \right\} \not=0 $ are found. Then the functions $f(x,l)$ and $g(x,l)$ are extracted and combined in one of the cases shown in Eqs.~(\ref{eq30})-(\ref{eq30b}). A single wavenumber is excited by taking  $f(x,l) = f(x) \delta(l - l_*)$ and $g(x,l) = g(x) \delta(l - l_*)$, where $\delta(l)$ is the Dirac delta function centered at wavenumber $l$. The value of $l$ is chosen to satisfy the periodic boundary conditions in the $y$-direction. Specifically, we take the value of $l = l_*$ shown in Fig.~\ref{f1}, corresponding to the third harmonic; the eigenfunctions were shown in Fig.~\ref{f_eigs}. The only effect the size of the transverse domain has on the linear stability is the quantization of the transverse wavenumbers. To satisfy $L_y$-periodic boundary conditions in $y$, the wavenumbers must be restricted to $k_n = 2 \pi n /L_y$ where $n \in \mathbb{Z}$. For these stationary modes we observe no dependence on the size of the transverse domain, unlike the type of transverse instability observed  in other 2D systems  \cite{Yamazaki,Klein3}. 
%%%%%%%%%%%%%%%%%%%%%%%%%%%%%%%%%%%%%%%%%%%%%%%%%%%%%%%%%%%%%%%%%%%%%%%%%%%%%%%%%%

A summary of the results from the simulations is %are 
shown in Fig.~\ref{evolution_snapshots}. For the reverse time solution in (\ref{eq17}), the forward time dynamics are the same as the classical NLS equation (\ref{eq16}).
Compared to the other equations, the PT  dynamics differ for different values of $\theta$  (see Fig.~\ref{evolution_snapshots}). When $\theta =0$, the PT evolution also resembles the other two. Three equally spaced filament pulses develop in all three cases, with the nonzero $\theta$ values shown developing faster.
From Figure \ref{evolution_snapshots}, it is evident for each case that a significant instability develops over the time intervals shown. Moreover, each case shows a neck-type instability, commonly observed in elliptic-focusing NLS-type systems \cite{Mamaev}. The spatial period of the bright spots match those predicted based on the wavenumber $l_*$ chosen. The distance between these  filaments is approximately $2\pi / l_* \approx 6.67$ units,
which is equal to the wavelength of the transverse perturbation.
%%%%%%%%%%%%%%%%%%%%%%%%%%%%%%%%%%%%%%%%%%%%%%%%%%%
\subsection{Stationary line soliton}
\begin{figure}
     \centering
     \includegraphics[scale=0.65]{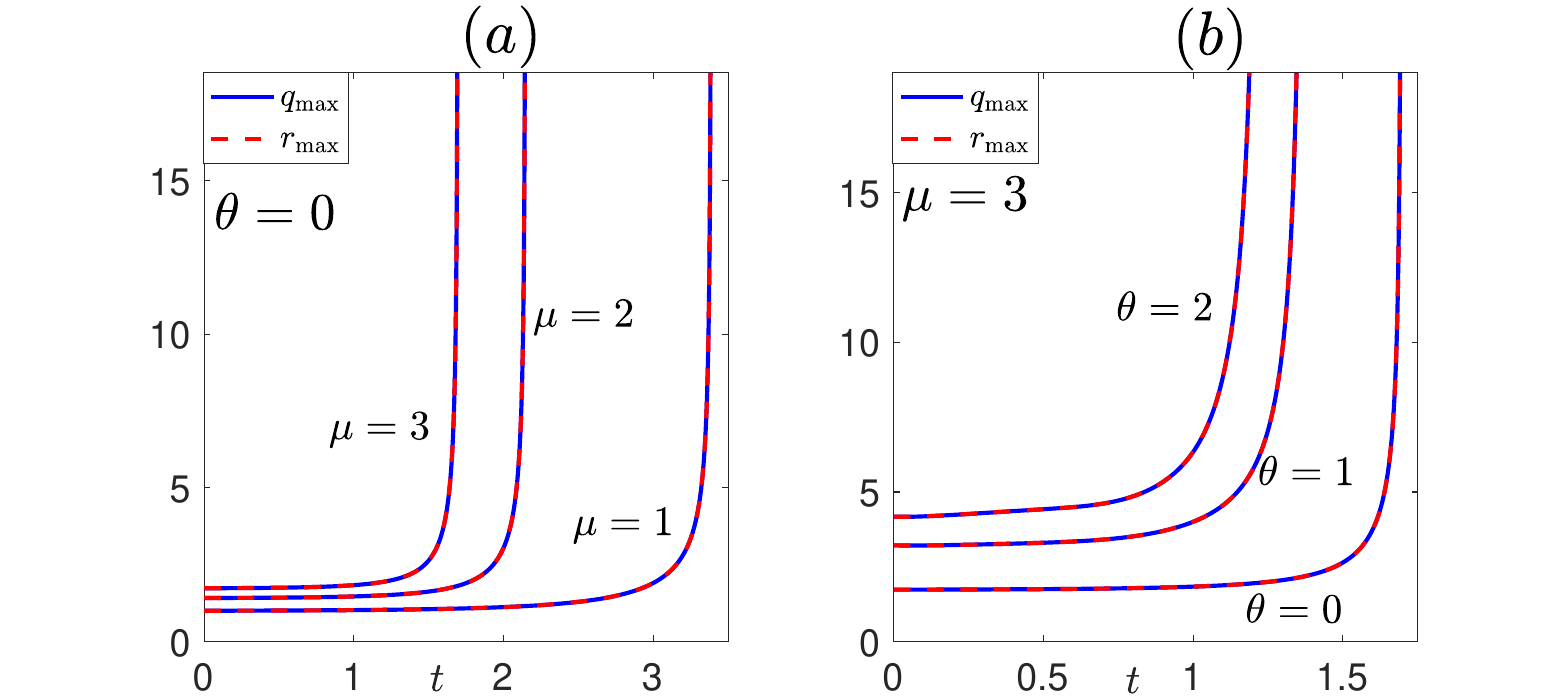}
\caption{Evolution of maximum magnitudes, $q_{\max}$ and $r_{\max}$ (\ref{eqqmax}), for different perturbed line solitons  (\ref{eq26})-(\ref{eq27}) in the PT equation (\ref{eq11}). Panel (a) varies $\mu$ for fixed $\theta = 0$, while panel (b) varies $\theta$ for fixed $\mu = 3$. The simulations in panel (b) correspond to those shown in Fig.~\ref{evolution_snapshots}. The $\theta = 0$ results are nearly  identical for all three equations.}  
        \label{f4}
\end{figure}

%%%%%%%%%%%%%%%%%%%%%%%%%%%%%%%%%%%%%%%%%%%%%%%%%%%
At larger times,  in all cases considered, we observe the peak magnitude growing,  apparently  without bound. This growth is reminiscent of the collapse in finite time observed in the 2D NLS equations with cubic nonlinearity \cite{Janssen,Vlasov,Sulem}. It is intriguing that the nonlocal NLS equations exhibit this feature too. %This in turn motivates the next section to analytically describe this apparent collapse.
In Fig.~\ref{evolution_snapshots} we have seen that a transversely perturbed line soliton for all cases leads to the formation of 2D filaments. The question we aim to address in this section is the long-time fate of these filaments vis-a-vis collapse.
To do so, we track the maximum amplitude %s in 
(\ref{eqqmax})
of an initially  perturbed line soliton solution.
For various soliton parameters, i.e. $\mu$ and $\theta$, (see Fig.~\ref{evolution_snapshots}) the maximum magnitude over the computational domain $\Omega$,
%%%%%%%%%%%%%%%%%%%%%%%%%%%%%%%%
is computed. The evolution of these quantities  for the %classical NLS and the 
PT NLS equation  (see (\ref{eq10})) is shown in Fig.~\ref{f4}. %We point out, however, that all three  reductions give similar results when $\theta = 0$. 
Here, for all parameters considered, the solutions appear to blow up (unbounded growth) in finite time. For fixed $\theta \not= \pi \left( \frac{1}{2} + n \right),$ the solitons generally become more susceptible to instability as $\mu$ grows, i.e. collapses faster. Additionally, for fixed $\mu$ this unbounded growth seems to  occur faster  as $\theta \rightarrow  \pi \left( \frac{1}{2} + n \right)$, where 
%$\theta \not= n \pi, n \in \mathbb{Z}$, where 
the PT solution (\ref{eq10}) moves away from the LOC solution. %reduces to a solution of the LOC equation (\ref{eq7}).
%%%%%%%%%%%%%%%%%%%%%%%%%%%%%%%%%%%%%%%%%%%%%%%%%%
%%%%%%%%%%%%%%%%%%%%%%%%%%%%%%%%%%%%%%%%%%%%%%%%%%%
%%%%%%%%%%%%%%%%%%%%%%%%%%%%%%%%%%%%%%%%%%%%%%%%%%%
\subsection{Quasi-breather line soliton}
%%%%%%%%%%%%%%%%%%%%%%%%%%%%%%%%%%%%%%%%%%%%%%%%%%%
Lastly, we consider the instability dynamics for the PT quasi-breather soliton \cite{AM13}
%%%%%%%%%%%%%%%
\begin{equation}
\label{breather_soln}
q_{\rm B}(x,t) = - \frac{2 (\eta + \overline{\eta})  }{ e^{ 4i \overline{\eta}^2 t + 2 \overline{\eta} x} + e^{4 i  \eta^2 t - 2 \eta  x} } , \;\;\; \eta, \overline{\eta} >0,
\end{equation}
%%%%%%%%%%%%%%%
which is localized in space and quasi-periodic in time whenever $\eta \not= \overline{\eta}$ (periodic in time when $\eta = \overline{\eta}$). The solution is called quasi-periodic since, in general, the frequencies $4\eta^2$ and $4 \overline{\eta}^2$ do not have commensurate periods in time. A singularity at $x = 0$ occurs when $\eta \not= \overline{\eta}$ and
$$t_s =  \frac{(2 n + 1)\pi}{4 (\eta^2 - \overline{\eta}^2)} , ~~~~~ n \in \mathbb{Z}.$$
The earliest positive time corresponds to $n = 0$ when $\eta > \overline{\eta}$. We can examine the growth rate of the breather at the singularity point $(x = 0)$. Here, $| e^{ 4i \overline{\eta}^2 t } + e^{4 i  \eta^2 t } | = \sqrt{2 + 2 \cos\left( 4 (\eta^2 - \overline{\eta}^2) t \right)}$ and near the singularity point $t_s$ %$t_s = \pi/(4 (\eta^2 - \overline{\eta}^2) )$ 
a Taylor expansion yields
\begin{align*}
 \sqrt{2 + 2 \cos\left( 4 (\eta^2 - \overline{\eta}^2) t \right)} %&% \approx \sqrt{2 + 2 \left( -1 + 16 (\eta^2 - \overline{\eta}^2)^2  \left[ t - t_s \right]^2 \right)} \\
 & =  4 (\eta^2 - \overline{\eta}^2)  \left[ t - t_s \right]  + O((t - t_s)^2).
 \end{align*}
Hence, as $t \rightarrow t_s$ the magnitude  approaches the singularity at $x= 0$ like 
 \begin{equation}
 \label{breather_growth}
 |q_B(0,t)| \approx \frac{1}{2 (\eta - \overline{\eta}) \left[ t - t_s \right]} \;, \;\;\;\; \eta \not= \overline{\eta}.
 \end{equation}

\begin{figure}
     \centering
     \includegraphics[scale=0.5]{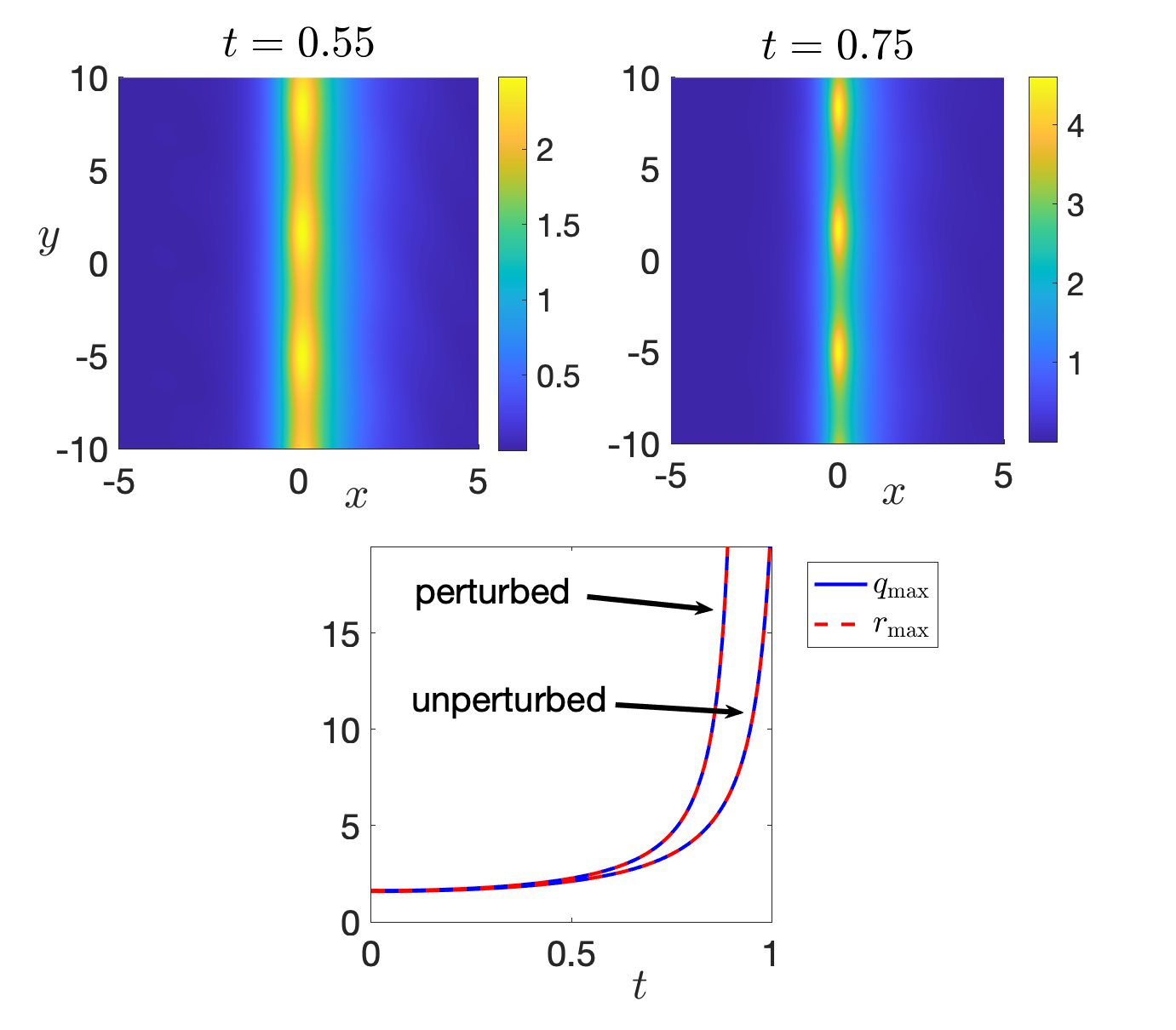}
\caption{(top) Snapshots of perturbed PT breather soliton $|q(x,y,t)|$ for $\eta = 1, \overline{\eta} = 0.5$ (see initial condition (\ref{breather_IC})) at indicated times. The quasi-breather soliton in (\ref{breather_soln}) develops a singularity at $t_s \approx 1.0 5$. (bottom) Peak amplitude evolutions (see (\ref{eqqmax})) for the perturbed ($\varepsilon = 0.5$) and unperturbed ($\varepsilon = 0$) initial conditions.}  
        \label{breather_instab_evolve}
\end{figure}

 The analysis of transverse instability of this type of solution does not fit within the framework developed in Sec.~\ref{linearstab}. 
 Rather, we examine it's stability properties through direct numerical simulations. To do so, we initially perturb (\ref{breather_soln}) by
 %%%%%%%%%%%%%%%%
 \begin{equation}
 \label{breather_IC}
q(x,y,0) = q_{\rm B}(x,0) + \varepsilon Q(x,y,0) ,
 \end{equation}
 %%%%%%%%%%%%%%%%
 with $Q(x,y,0)$ given by (\ref{eq30b}) using the eigenfunctions shown in Fig.~\ref{f_eigs} (second row). The numerical results found by solving the $q,r$ system (\ref{eq1})-(\ref{eq2}) and PT reduction (\ref{eq10}) are shown in Fig.~\ref{breather_instab_evolve}. The soliton is observed to develop 2D filaments, similar to those observed in Fig.~\ref{evolution_snapshots} for bounded solutions. Hence, the ``neck'' type instability development of line solitons nonlocal reductions appears to be a rather generic feature.
For the type of solution %perturbation 
considered here, both the perturbed and unperturbed quasi-breather solitons develop singularities in finite time. %Here, 
However, the singularity formation of the perturbed mode occurs earlier than the unperturbed quasi-breather. Hence, the collapse process of an already singular solution appears to accelerate in the presence of a transverse pertubration (see bottom Fig.~\ref{breather_instab_evolve}). 
%%%%%%%%%%%%%%%%%%%%%%%%%%%%%%%%%%%%%%%%%%%%%%%%%%
%%%%%%%%%%%%%%%%%%%%%%%%%%%%%%%%%%%%%%%%%%%%%%%%%%
\section{Conclusions}
\label{conclude}
%%%%%%%%%%%%%%%%%%%%%%%%%%%%%%%%%%%%%%%%%%%%%%%%%%
%%%%%%%%%%%%%%%%%%%%%%%%%%%%%%%%%%%%%%%%%%%%%%%%%%
%%%%%%%%%%%%%%%%%%%%%%%%%%%%%%%%%%%%%%%%%%%%%%%%%%
In this paper we have provided an extensive numerical study exploring collapse and blow-up in finite time for the two-dimensional space-time nonlocal nonlinear Schr\"odinger system. Using a Gaussian initial condition centered at $(x_0,y_0)$, we were able to identify a regime in parameter space where collapse could occur. A so-called quasi-variance quantity was introduced in an attempt to describe the dynamics of wave collapse. It turns out that for the nonlocal PT equation, this quantity could not be used to predict collapse in finite time. To shed more light on the issue of finite time blow-up, we showed that line soliton solutions of nonlocal NLS variants are transversely unstable subject to long wavelength modulations. The stability profiles of the nonlocal variants resembled those of the classical (local) NLS equation. Motivated by numerics, we also studied blow-up in line soliton solutions.  

There are still several more nonlocal variants of the two-dimensional extensions associated with the integrable (1+1)D Ablowitz-Musslimani systems that are important to study. It would be interesting to see if the stability and existence characteristics persist in other integrable, but nonlocal NLS equations, or perhaps some new twists will be discovered. Furthermore, we have only studied effectively simple class of solutions of these equations. It would be interesting to study stability properties and collapse of more general solutions (beyond the one presented here) of these nonlocal systems.
Finally, an even more general question is, what are the stability/collapse properties of the general two-dimensional $q,r$ system, away from symmetry reductions. %points. 
%%%%%%%%%%%%%%%%%%%%%%%%%%%%%%%%%%%%%%%%%%%%%%%%%%%%%%%%%%%%%%%%%%%%%%%%%%%%%%%
\vspace{1 in}

%%%%%%%%%%%%%%%%%%%%%%%%%%%%%%%%%%%%%%%%%%%%%%%%%%%%%%%%%%%%%%%%%%%%%%%%%%%%%%%

\end{document}